\documentclass[a4paper,11pt]{article}
\pdfoutput=1 % if your are submitting a pdflatex (i.e. if you have
\usepackage{jcappub}
\usepackage[utf8]{inputenc}
\usepackage[T1]{fontenc} % if needed
\usepackage{physics} % For e.g. \qty()
\newcommand{\Mpl}{M_\mathrm{Pl}}

\usepackage{natbib}

\newcommand{\bea}{\begin{eqnarray}} \newcommand{\eea}{\end{eqnarray}}
\newcommand{\reh}{{\mathrm{reh}}}
\newcommand{\el}{\nonumber \\}
\newcommand{\re}[1]{(\ref{#1})}

\title{Higgs inflation at the hilltop}

\author{Vera-Maria~Enckell,}
\author{Kari~Enqvist,}
\author{Syksy~R{\"a}s{\"a}nen}
\author{and Eemeli~Tomberg}
\affiliation{University of Helsinki, Department of Physics and Helsinki Institute of Physics,\\ P.O. Box 64, FIN-00014 University of Helsinki, Finland}

\emailAdd{vera-maria.enckell@helsinki.fi}
\emailAdd{kari.enqvist@helsinki.fi}
\emailAdd{syksy.rasanen@iki.fi}
\emailAdd{eemeli.tomberg@helsinki.fi}

\abstract{
We study inflation with the non-minimally coupled Standard Model Higgs in the case when quantum corrections generate a hilltop in the potential. We consider both the metric and the Palatini formulation of general relativity. We investigate hilltop inflation in different parts of the Higgs potential and calculate predictions for CMB observables. We run the renormalization group equations up from the electroweak scale and down from the hilltop, adding a jump in-between to account for unknown corrections in the intermediate regime. Within our approximation, no viable hilltop inflation is possible for small field values, where the non-minimal coupling has no role, nor for intermediate field values. For large field values, hilltop inflation works. We find the spectral index to be $n_s\leq0.96$ in both the metric and the Palatini formulation, the upper bound coinciding with the tree-level result. The tensor-to-scalar ratio is $r\leq1.2\times10^{-3}$ in the metric case and $r\leq2.2\times10^{-9}$ in the Palatini case.
Successful inflation is possible even when the renormalization group running is continuous with no jumps.
In the metric formulation, $r$ is smaller than in Higgs inflation on the tree-level plateau or at the critical point, making it possible to distinguish hilltop inflation from these scenarios with next-generation CMB experiments.
}

\begin{document}

\begin{flushleft}
	\hfill		 HIP-2018-8/TH \\
\end{flushleft}

\maketitle
\flushbottom

\section{Introduction} \label{sec:intro}

\paragraph{Extending the Standard Model to high scales and Higgs inflation.}

No deviations from the Standard Model (SM) of particle physics have been discovered in collider experiments. The only new piece of information (stronger exclusion limits on extensions of the SM aside) uncovered by LHC is the Higgs mass, $m_H=125.09\pm0.21\pm0.11$ GeV \cite{Aad:2015zhl}. This value is such that the SM may be valid for Higgs field values up to the Planck scale $\Mpl=2.4\times10^{18}$ GeV and possibly beyond. However, it is also possible that top mass loop contributions drive the quartic Higgs self-coupling negative at large field values, making the electroweak vacuum unstable. The stability limit is sensitive to the precise values of the Higgs and top mass and the strong coupling constant. Present values are consistent with stability, instability, or metastability at the Planck scale, within the experimental and theoretical uncertainties \cite{Espinosa:2015a, Espinosa:2015b, Iacobellis:2016, Butenschoen:2016lpz, Espinosa:2016nld}. Validity of the SM up to high scales means that the Higgs field could be the inflaton \cite{Bezrukov:2007ep, Bezrukov:2013fka}.

Inflation is the most successful scenario for the primordial universe. It alleviates the homogeneity and isotropy problem \cite{Vachaspati:1998dy, Trodden:1999wc, Ellis:1999sx, ellis2000, East:2015ggf}, explains spatial flatness and has (in its simplest variants) predicted in detail that initial perturbations are mostly adiabatic, close to scale-invariant, highly Gaussian and predominantly scalar \cite{Starobinsky:1980te, Kazanas:1980tx, Guth:1980zm, Sato:1980yn, Mukhanov:1981xt, Linde:1981mu, Albrecht:1982wi, Hawking:1981fz, Chibisov:1982nx, Hawking:1982cz, Guth:1982ec, Starobinsky:1982ee, Sasaki:1986hm, Mukhanov:1988jd}. This is in excellent agreement with observations \cite{Ade:2015lrj}. If the non-minimal coupling of the Higgs field to gravity is neglected, then at large field values the potential can be approximated as $\frac{1}{4}\lambda(\phi)\phi^4$, where the coupling $\lambda(\phi)$ depends on the field due to loop corrections. This potential is not sufficiently flat to allow a long enough period of inflation with a small enough amplitude of perturbations \cite{Isidori:2007vm, Hamada:2013mya, Fairbairn:2014nxa}.

As is now well known, once the Higgs is non-minimally coupled to gravity, the inflationary behaviour changes. Even if the non-minimal coupling $\xi$ is put to zero at the classical level, it will be generated by quantum corrections \cite{Callan:1970ze}.
Inflation with the non-minimal coupling has been considered since the 1980s \cite{Futamase:1987ua, Salopek:1988qh, Fakir:1990eg, Makino:1991sg, Fakir:1992cg, Kaiser:1994vs, Komatsu:1999mt}, and Higgs inflation is a remarkably simple realization. It uses the only known scalar field that may be elementary instead of a composite, and does not introduce any new parameters that are not required by the theory.
If only the tree-level SM Higgs potential is considered, the non-minimal coupling makes the potential flat at large field values, and the inflationary predictions for the spectral index and the tensor-to-scalar ratio only depend on the number of e-folds until the end of inflation \cite{Bezrukov:2007ep}. As the SM field content and couplings are precisely known, preheating and reheating can be calculated in detail, removing the uncertainty in the number of e-folds that accompanies less complete embeddings of inflation into particle physics \cite{Bezrukov:2008ut, GarciaBellido:2008ab, Figueroa:2015, Repond:2016} (although see also \cite{Ema:2016}), with the caveat that there can be modifications due to physics beyond the SM. The tree-level predictions agree well with observations \cite{Ade:2015lrj}. Higgs decay also produces a distinctive signature of gravitational waves (also present, though different, when Higgs is not the inflaton) \cite{Figueroa:2014aya}.

Loop corrections complicate the picture. The quantum correction to the potential, and thus the inflationary predictions, depend on the Higgs and top masses measured at low energies. In principle this offers a novel consistency test between collider experiments and cosmological observations \cite{Espinosa:2007qp, Barvinsky:2008ia, Burgess:2009ea, Popa:2010xc, DeSimone:2008ei, Bezrukov:2008ej, Bezrukov:2009db, Barvinsky:2009ii, Bezrukov:2010jz, Bezrukov:2012sa, Allison:2013uaa, Salvio:2013rja, Shaposhnikov:2013ira}.
However, as the SM coupled to gravity is non-renormalizable, it is not clear how the loop corrections should be calculated, and the relation between the low and high energy regimes is still debated \cite{George:2013iia, Calmet:2013hia, Bezrukov:2014bra, Bezrukov:2014ipa, Rubio:2015zia, George:2015nza, Saltas:2015vsc, Fumagalli:2016lls, Enckell:2016xse, Moss:2014, Bezrukov:2017dyv}. (Possible violation of perturbative unitarity is another issue that remains unsettled \cite{Barbon:2009ya, Burgess:2009ea, Hertzberg:2010dc, Bauer:2010jg, Bezrukov:2010jz, Bezrukov:2011sz, Calmet:2013hia, Weenink:2010rr, Prokopec:2012ug, Xianyu:2013rya, Prokopec:2014iya, Ren:2014sya, Escriva:2016cwl}.)
At low energies, where gravity can be neglected, the SM is renormalizable, and at high energies there is an approximate shift symmetry that keeps loop corrections under control, but the matching between the two regimes is not uniquely specified \cite{Bezrukov:2010jz, George:2013iia, Bezrukov:2014ipa, George:2015nza, Fumagalli:2016lls, Enckell:2016xse, Bezrukov:2017dyv}.

Quantum corrections do not only affect the mapping between observables at the electroweak scale and on the inflationary plateau. They can also open new inflationary regimes by qualitatively changing the shape of the potential. In addition to inflation on the plateau as originally proposed, it may be possible for Higgs inflation to occur at a near-inflection point (called a critical point) \cite{Allison:2013uaa, Bezrukov:2014bra, Hamada:2014iga, Bezrukov:2014ipa, Rubio:2015zia, Fumagalli:2016lls, Enckell:2016xse, Bezrukov:2017dyv, Rasanen:2017ivk}, at a hilltop \cite{Fumagalli:2016lls, Rasanen:2017ivk} or when climbing up a hill from a degenerate vacuum \cite{Jinno:2017a, Jinno:2017b}.\footnote{It is also possible to have inflation in a false vacuum, although new physics is needed for graceful exit to the electroweak vacuum \cite{Masina:2011aa, Masina:2011un, Masina:2012yd, Masina:2014yga, Notari:2014noa, Fairbairn:2014nxa, Iacobellis:2016}.}

We present the first detailed study of Higgs inflation at the hilltop that uses renormalization group equations to connect the electroweak and inflationary scales. Starting at a hilltop requires tuning, so it is easily missed when the parameter space is scanned starting from the electroweak scale, as the shape of the potential is sensitive to small changes in the observed Higgs and top masses. In contrast, we start from the assumption that there is a hilltop, and use both the existence of the hilltop and the measured electroweak scale observables as input. We consider both the metric and the Palatini formulation of general relativity \cite{einstein1925, ferraris1982}. When the Higgs is non-minimally coupled, they represent two distinct physical theories \cite{Bauer:2008zj, Bauer:2010jg, Rasanen:2017ivk}.

In section \ref{sec:higgsinfl} we introduce the action in the metric and the Palatini formulation, explain how we calculate the quantum corrections and introduce the equations of motion and the observables. In section \ref{sec:small_and_interm} we show that within our approximation, hilltop inflation is ruled out at small field values, where the non-minimal coupling does not play a role. While hilltop inflation may be possible in the intermediate regime, within our approximation the spectrum would not agree with observations. In \ref{sec:largefield} we discuss analytically why successful hilltop inflation is possible at large field values, and numerically scan the parameter space. We contrast the metric and the Palatini formulation and compare to previous work. In particular, we find that in the metric case the tensor-to-scalar ratio at the hilltop is always smaller than on the tree-level plateau or at the critical point, while this is not true in the Palatini formulation. We summarise our findings in section \ref{sec:conclusions}.

\section{Non-minimally coupled Higgs inflation} \label{sec:higgsinfl}

\subsection{Tree-level action}

In Higgs inflation, the action is
\begin{equation} \label{eq:higgs_SM_action}
	S = \int d^4 x \sqrt{-g} \qty[ -\frac{1}{2}\left(M^2 + \xi h^2\right) g^{\mu\nu} R_{\mu\nu} - \frac{1}{2} g^{\mu\nu} \partial_\mu h \partial_\nu h - V(h) + \mathcal{L}_{SM} ] \ ,
\end{equation}
where $M$ is a mass scale (close to the Planck scale) which we set to unity henceforth, $\xi$ is the non-minimal coupling, $h$ is the radial Higgs field, $g_{\mu\nu}$ is the metric, $R_{\mu\nu}$ is the Ricci tensor, and $\mathcal{L}_{\mathrm{SM}}$ includes the rest of the SM.

To obtain the equations of motion, the action \eqref{eq:higgs_SM_action} is minimized with respect to the independent degrees of freedom. In the case of gravity, there are several ways of choosing what they are. One possibility is the metric formulation, where it is imposed from the beginning that the connection is Levi--Civita, so that it is written in terms of the metric (and its derivatives), which is the only gravitational degree of freedom. In this case, the York--Gibbons--Hawking boundary term \cite{York:1972, Gibbons:1977} has to be added to the action to cancel a boundary term in the variation.

Another possibility is the Palatini formulation, where the metric and the connection are taken to be independent variables. In this case, the Ricci tensor depends only on the connection (and its first derivative), not on the metric nor its derivatives. For the Einstein--Hilbert action plus an action for matter that couples only to the metric, not to its derivatives nor to the connection, minimizing the action with respect to the connection then leads to the Levi--Civita connection \cite{einstein1925, ferraris1982} and thus to the same physics as in the metric case. The Palatini formulation is simpler in the sense that it involves less assumptions, and there is no need to add a boundary term to obtain the equations of motion. When matter couples to derivatives of the metric or to the connection (or the gravitational action is modified), in particular when there is a non-minimally coupled scalar field, these two formulations are different physical theories \cite{Bauer:2008zj, Bauer:2010jg}. We will study both of them.

On inflationary scales, the Higgs mass term is negligible, so we have
\begin{equation} \label{eq:higgs_pot}
	V(h) = \frac{\lambda}{4}h^4 \ ,
\end{equation}
where $\lambda$ is the Higgs self-coupling. It is convenient to make the conformal transformation $g_{\alpha\beta}\rightarrow (1+\xi h^2)^{-1} g_{\alpha\beta}$ to the Einstein frame where the scalar field is minimally coupled\footnote{Classically and at tree-level, the physics is independent of the frame \cite{Chiba:2013mha, Postma:2014vaa, Jarv:2016sow, Karamitsos:2017elm, Karamitsos:2018lur, Makino:1991sg, Fakir:1992cg, Komatsu:1999mt, Koh:2005qp, Chiba:2008ia, Weenink:2010rr, Kubota:2011, Flanagan:2004bz, Jarv:2014hma, Jarv:2015kga, Kuusk:2015dda, Ren:2014sya}. For loop corrections, the issue is less clear \cite{Weenink:2010rr, Calmet:2012eq, Prokopec:2014iya, Kamenshchik:2014waa, Burns:2016ric, Fumagalli:2016lls, Hamada:2016, Karamitsos:2017elm, Karamitsos:2018lur, Bezrukov:2008ej, Bezrukov:2009db, Bezrukov:2010jz, Allison:2013uaa, George:2013iia, Postma:2014vaa, Prokopec:2012ug, Herrero-Valea:2016jzz, Pandey:2016jmv, Pandey:2016unk}.}. A field $\chi$ with a canonical kinetic term is recovered with the field transformation
\begin{equation} \label{eq:dchi_dh}
	\frac{d h}{d \chi} = \frac{1 + \xi h^2}{\sqrt{1 + \xi h^2  + p6\xi^2 h^2}} \ .
\end{equation}
Here $p=1$ in the metric formulation and $p=0$ in the Palatini formulation \cite{Bauer:2008zj}. This difference is a direct consequence of the choice of degrees of freedom. As the conformal transformation only changes the metric, in the Palatini case the Ricci tensor is left invariant, which results in a different kinetic term in the Einstein frame. In general, the difference between the metric and the Palatini formulation can be shifted between the connection, scalar field kinetic term and potential by field and conformal transformations, which do not change the physical content of the theory.

Now the inflationary behaviour is determined by the potential as a function of $\chi$ in the Einstein frame. It takes the form \cite{Bezrukov:2007ep}
\begin{equation} \label{eq:higgs_pot_einstein}
	U(\chi) \equiv \frac{V[h(\chi)]}{[1+\xi h(\chi)^2]^2} \equiv
	 \frac{\lambda}{4}F[h(\chi)]^4 \ , \qquad F(h) \equiv \frac{h}{\sqrt{1+\xi h^2}} \ .
\end{equation}
In the metric formulation, the function $F$ can be approximated in terms of $\chi$ as \cite{Bezrukov:2007ep}
\begin{equation} \label{eq:F_approx}
	F(\chi) \approx \begin{cases}
    \chi & \qquad h \ll 1/\xi \\
    \frac{1}{\sqrt{\xi}}\qty(1-e^{-\sqrt{2/3}\chi})^{1/2} & \qquad h \gg 1/\sqrt{\xi} \ .
  \end{cases}
\end{equation}
There are three regimes for the field: small values $h \ll 1/\xi$, large values $ h \gg 1/\sqrt{\xi}$, and intermediate values in between. In particular, for large field values, the potential has a plateau where slow-roll inflation is possible.

In the Palatini formulation, the relation \eqref{eq:dchi_dh} can be solved exactly, with the result that
\begin{equation} \label{eq:F_palatini}
	F(\chi) = \frac{1}{\sqrt{\xi}}\tanh \qty(\sqrt{\xi}\chi) \ .
\end{equation}
Again, there is a plateau at large field values.

\subsection{Loop corrected effective potential} \label{sec:quantumcorrs}

The above considerations are based on classical physics. We can take quantum corrections into account by replacing the classical potential with the loop-corrected potential. We do this in the Einstein frame and calculate the effective potential in flat spacetime, using the $\overline{MS}$ renormalization scheme.

The problem with quantum corrections is that the model is not renormalizable. However, for small field values $h \ll 1/\xi$ it can be approximated by the SM without non-minimal coupling. In the small field limit we can thus calculate the loop corrections in a straightforward way. For $h \gg 1/\sqrt{\xi}$, the function $F$ defined in \eqref{eq:higgs_pot_einstein} approaches a constant, and as a consequence the Higgs field becomes massless and decouples from the other fields. The Higgs then no longer plays a dynamical role in the theory, and we are left with the chiral electroweak theory, also called the chiral SM \cite{Bezrukov:2009db}. While the chiral SM is not renormalizable, it can be treated consistently order by order, and the leading correction is calculable. The connection between these two regions is not uniquely determined within the model, but can be parametrised with effective jumps in couplings at the matching \cite{Bezrukov:2014bra,Bezrukov:2014ipa}.

For small field values, we use the SM two-loop effective potential \cite{Ford:1992pn}. We calculate it with the code available at \cite{SMEffPotCode}, see also \cite{Martin:2003qz, Martin:2005qm, Martin:2014cxa}. We run the parameters using the three-loop beta functions given in the code available at \cite{SMRunningCode}, see also \cite{Chetyrkin:2012rz, Bezrukov:2012sa}. We include masses for the Higgs radial mode, the would-be Goldstone bosons, the $W$ and $Z$ bosons and the top quark; the other fermions are treated as massless. The code is used to determine the running and the initial values for the couplings, starting from the observed values\footnote{The code \cite{SMRunningCode} also uses as fixed initial values the $W$ and $Z$ boson pole masses $m_W=80.399$ GeV and $m_Z=91.1876$ GeV, the Fermi constant $G_F=1.16637 \times 10^{-5}$ GeV\textsuperscript{$-2$}, the fine structure constant $\alpha=1/127.916$ and $\sin^2 \theta_W=0.23116$, where $\theta_W$ is the Weinberg angle, both evaluated at $m_Z$ in the $\overline{MS}$ scheme. These fix the initial values of the electroweak gauge couplings $g$ and $g'$.}
\begin{equation} \label{eq:SM_bestfit_vals}
	\frac{g_S^2(m_Z)}{4\pi}=0.1184 \ , \quad m_H = 125.09 \pm 0.24\,\mathrm{GeV} \ , \quad m_t = 172.44 \pm 0.49\,\mathrm{GeV} \ .
\end{equation}
We quote error bars as 68\% confidence intervals and limits as 95\% confidence intervals.
Here $g_S(m_Z)$ is the strong coupling at the scale of the $Z$ boson mass $m_Z$, and $m_H$ and $m_t$ are the pole masses of the Higgs boson and the top quark \cite{Aad:2015zhl, Khachatryan:2015hba}. The above error bars for the top quark do not include a systematic error from matching the theoretical pole mass (relevant for renormalization group running) to the observationally determined MCMC mass, quoted as being of the order 1 GeV in \cite{Khachatryan:2015hba} and estimated as between 0.2 and 0.4 GeV in \cite{Butenschoen:2016lpz}. We combine the estimate 0.4 GeV quadratically with the error given in \re{eq:SM_bestfit_vals} for a total 1$\sigma$ error of 0.6 GeV.
We run the couplings up to the scale that equals the background Higgs field value:
\begin{equation} \label{eq:renormscale_SM}
	\mu(h) = h \ .
\end{equation}
For large field values we have the chiral SM, and there we add to the tree-level potential \eqref{eq:higgs_pot_einstein} only the one-loop correction, which reads \cite{Bezrukov:2009db}
\begin{equation} \label{eq:large_potential_corr}
	U_{\mathrm{1-loop}} = \frac{6m_W^4}{64\pi^2}\left(\ln \frac{m_W^2}{\mu^2} - \frac{5}{6} \right) +
				\frac{3m_Z^4}{64\pi^2}\left(\ln \frac{m_Z^2}{\mu^2} - \frac{5}{6} \right) -
				\frac{3m_t^4}{16\pi^2}\left(\ln \frac{m_t^2}{\mu^2} - \frac{3}{2} \right) \ ,
\end{equation}
where the $W$ boson, $Z$ boson and top quark masses are given by
\begin{equation} \label{eq:large_masses}
	m_W^2 = \frac{g^2 F^2}{4} \ , \qquad m_Z^2 = \frac{\qty(g^2+g'^2) F^2}{4} \ , \qquad m_t^2 = \frac{y_t^2 F^2}{2} \ ,
\end{equation}
where $y_t$ is the top Yukawa coupling and $g$ and $g'$ are the electroweak gauge couplings.
For the chiral SM, we use the one-loop beta functions, which read \cite{Bezrukov:2009db, Dutta:2007st}
\bea \label{eq:large_betas}
%\begin{split}
	16\pi^2\beta_\lambda &=& -6y_t^4 + \frac{3}{8}\qty[2g^4 + (g'^2+g^2)^2] \ , \quad 16\pi^2\beta_{y_t} = y_t\qty(-\frac{17}{12}g'^2 - \frac{3}{2}g^2 - 8g_S^2 + 3y_t^2) \ , \el
	16\pi^2\beta_{g} &=& -\frac{13}{4}g^3 \ , \quad
	16\pi^2\beta_{g'} = \frac{27}{4}g'^3 \ , \quad
	16\pi^2\beta_{g_S} = -7g_S^3 \ .
%\end{split}
\eea
We have dropped from $\beta_\lambda$ a term that is proportional to $\lambda$, since it is a higher order correction near a hilltop, where $\beta$ is of order $\lambda$ (see section \ref{sec:large_field_anal}). We neglect the running of the non-minimal coupling $\xi$.

There is no difference between the metric and the Palatini formulation for small field values. However, in the large field regime the renormalization group running will be different \cite{Markkanen:2017tun}, because the Einstein frame kinetic term is different. (Note that it is not possible to obtain both minimal coupling and a canonical kinetic term for all fields in the Higgs doublet \cite{Hertzberg:2010dc, Mooij:2011fi, Greenwood:2012aj, George:2013iia, George:2015nza}.) We neglect this difference.

For the large field case we choose the renormalization scale to be
\begin{equation} \label{eq:renormscale_chiral}
	\mu(\chi) = \gamma F(\chi) \ ,
\end{equation}
chosen so that the explicit $F(\chi)$-dependence of the masses cancels in the one-loop correction \eqref{eq:large_potential_corr}. The constant $\gamma$ is determined from the condition that the correction vanishes at the hilltop.

To match the two regions, we switch from the SM running to the chiral SM running \eqref{eq:large_betas} at the scale
\begin{equation} \label{eq:jumpscale}
	\mu_1 = \frac{\gamma}{\xi} \ ,
\end{equation}
chosen so that it is below the large field regime, but well above the electroweak scale. At this scale, we let the couplings $\lambda$ and $y_t$ jump {by $\Delta\lambda$ and $\Delta y_t$, respectively. These jumps parametrise the effect of unknown physics in the region $1/\xi \lesssim h \lesssim 1/\sqrt{\xi}$. For simplicity, we neglect such jumps in the gauge couplings $g_S$, $g$ and $g'$. The jump parameters $\Delta\lambda$ and $\Delta y_t$ are also assumed not to run.

As the theory is non-renormalizable, new kinds of contributions arise at every loop order, and may be important in the matching region. We will not consider them in any detail nor explicitly introduce any new couplings in the renormalization group equations. However, in addition to the instant jumps, we consider the transition from the SM to the chiral SM in more detail in section \ref{sec:intermediate} by postulating a smooth transition instead of instant jumps in $\lambda$ and $y_t$:
\begin{equation} \label{eq:lambda_jump}
\begin{gathered}
	\lambda(\chi) = \lambda_0 + \Delta\lambda \, S(\chi) \ , \\
	S(\chi) \equiv \left\{ \left[F'(\chi)\right]^2 + \frac{1}{3}F''(\chi)F(\chi) \right\}^2 - 1 \ ,
\end{gathered}
\end{equation}
motivated by theoretical considerations \cite{Bezrukov:2014ipa}. The step-function-like $S(\chi)$ behaves as
\begin{equation}
	\label{eq:step_limits}
	S(\chi) \xrightarrow{h \ll 1/\xi} 0 \ , \qquad
	S(\chi) \xrightarrow{h \gg 1/\sqrt{\xi}} -1 \ .
\end{equation}
We keep $\lambda_0$ and $\Delta\lambda$ constant and approximate the effective potential as
\begin{equation} \label{eq:interm_pot}
	U(\chi)=\frac{\lambda(\chi)}{4}F(\chi)^4 \ ,
\end{equation}
which applies in the region $1/\xi \lesssim h \lesssim 1/\sqrt{\xi}$.

In most cases, the loop corrections do not significantly change the shape of the tree-level potential, they simply reduce the value of $\lambda$ for large fields \cite{Fumagalli:2016lls, Fumagalli:2016sof}. However, for certain fine-tuned parameter values, the corrections may cause a qualitative change in the potential, generating an inflection point \cite{Allison:2013uaa, Bezrukov:2014bra, Hamada:2014iga, Bezrukov:2014ipa, Rubio:2015zia, Fumagalli:2016lls, Enckell:2016xse, Rasanen:2017ivk}, a hilltop \cite{Fumagalli:2016lls, Rasanen:2017ivk} or a degenerate vacuum \cite{Jinno:2017a, Jinno:2017b}.

\subsection{Equations of motion and inflationary observables} \label{sec:slowroll}

In the spatially flat Friedmann--Robertson--Walker universe, the Einstein equation gives
\begin{equation}
	\label{eq:friedmann_scalar_eq}
	3 H^2 = \frac{1}{2}\dot{\chi}^2 + U(\chi) \ , \qquad
	\ddot{\chi} + 3H\dot{\chi} + U'(\chi) = 0 \ ,
\end{equation}
where $H$ is the Hubble parameter, dot denotes derivative with respect to cosmic time $t$ and prime denotes derivative with respect to $\chi$. Slow-roll inflation can only occur for field values for which the slow-roll parameters characterising the potential $U(\chi)$ are small. The first order slow-roll parameters are
\begin{equation}
	\label{eq:sr_1st_order}
	\epsilon \equiv \frac{1}{2} \qty(\frac{U'}{U})^2 \ , \qquad
	\eta \equiv \frac{U''}{U} \ ,
\end{equation}
and in addition there is a series of higher order parameters \cite{Liddle:1994}:
\begin{align}
	\label{eq:sr_higher_order}
	\zeta \equiv \frac{U'}{U} \frac{U'''}{U} \ , \qquad \varpi \equiv \qty(\frac{U'}{U})^2 \frac{U''''}{U}  \ , \qquad \dots
\end{align}
During slow-roll, the number of e-folds between scale factor $a$ and the end of inflation, when $a=a_\mathrm{end}$, is
\begin{equation} \label{eq:efolds}
	N \equiv \ln\frac{a_\mathrm{end}}{a} = \int_{\chi_\mathrm{end}}^{\chi} \frac{1}{\sqrt{2\epsilon}} d\chi \ .
\end{equation}
The number of e-folds at the pivot scale $k_*=0.05$ Mpc$^{-1}$ is
\begin{equation} \label{Npivot}
\begin{split}
  N_* &= 61 - \Delta N_\reh + \frac{1}{4}\ln U_* + \frac{1}{4} \ln\frac{U_*}{U_\mathrm{end}} \\
  &= 52 - \frac{1}{4} \ln\frac{0.09}{r_*} \ ,
\end{split}
\end{equation}
where the subscript end refers to the end of inflation and $\Delta N_\reh$ is the number of e-folds between the end of inflation and the end of preheating (defined as the time when energy density starts to scale like radiation). On the second line we have taken into account that for SM field content $\Delta N_\reh=4$ \cite{Figueroa:2009,Figueroa:2015,Repond:2016} (although see \cite{Ema:2016}), written $U_*=\frac{3\pi^2}{2} A_s r$ and inserted the observed value $24\pi^2 A_s=5.2\times10^{-7}$ \cite{Ade:2015lrj}, written in the maximum value 0.09 of $r_*$ allowed by CMB observations as a point of comparison, and taken into account that the term describing the change of the potential between the pivot scale and the end of inflation is $<1$.

The spectral index $n_s$, tensor-to-scalar ratio $r$, running of the spectral index $\alpha_s$, and running of the running $\beta_s$ are determined, to lowest order, in terms of the slow-roll variables as \cite{Planck:2013jfk}:
\begin{equation} \label{eq:sr_parameters}
\begin{split}
	n_s &= 1-6\epsilon+2\eta \ , \qquad \qquad \ r = 16 \epsilon \ , \\
	\alpha_s &= 16\epsilon\eta - 24\epsilon^2 - 2\zeta \ , \qquad
	\beta_s = -192\epsilon^3 + 192\epsilon^2\eta - 32\epsilon\eta^2 - 24\epsilon\zeta + 2\eta\zeta + 2\varpi \ .
\end{split}
\end{equation}
When both $\alpha_s$ and $\beta_s$ are allowed to vary, at the pivot scale Planck temperature and low multipole polarization data give \cite{Ade:2015lrj} %(see equation (19) in \cite{Ade:2015lrj})
\begin{equation}
	\label{eq:sr_obs_n}
	n_s=0.9569\pm0.0077 \ , \quad \alpha_s=0.011^{+0.014}_{-0.013} \ , \quad \beta_s=0.029^{+0.015}_{-0.016} \ ,
\end{equation}
while combining with BICEP2/Keck data gives \cite{r}
\begin{equation}
	\label{eq:sr_obs_r}
	r<0.09 \ .
\end{equation}
We already used the normalization of the scalar perturbations at the pivot scale, %\cite{Ade:2015lrj}
\begin{equation}
	\label{eq:pert_norm}
	\frac{U}{\epsilon} = 5.2 \times 10^{-7} \ .
\end{equation}
Tree-level Higgs inflation on the plateau predicts \cite{Bezrukov:2007ep}
\bea \label{eq:higgsinfl_standard_results}
	n_s = 1 - \frac{2}{N} \ , \qquad r = \frac{12}{N^2} \ , \qquad
	\alpha_s = -\frac{2}{N^2} \ , \qquad \beta_s = -\frac{4}{N^3} \ .
\eea
For $N=50$ we get
\begin{equation} \label{eq:higgsinfl_standard_results_N50}
	n_s = 0.96 \ , \qquad r = 4.8 \times 10^{-3} \ , \qquad \alpha_s = - 0.8 \times 10^{-3} \ , \qquad \beta_s = - 3.2 \times 10^{-5}
\end{equation}
in good agreement with the observational values \eqref{eq:sr_obs_n} and \re{eq:sr_obs_r}.

\subsection{Hilltop inflation} \label{sec:hilltop}

Successful models of single-field inflation must satisfy \eqref{eq:sr_obs_n}, \eqref{eq:sr_obs_r} and \eqref{eq:pert_norm}, together with \re{Npivot}. One class of models is those where the potential has a local maximum \cite{Boubekeur:2005zm, Kohri:2007gq}. At the hilltop, the slow-roll parameter $\epsilon$ is zero, and if the potential is flat enough in its vicinity, the other slow-roll parameters can also be small. Then it is possible to have slow-roll inflation near the hilltop.

Because the number of e-folds of inflation approaches infinity as the initial value of the inflaton approaches the hilltop, it is easy to have a long duration of inflation. If the field starts close enough to the maximum, its evolution will be dominated by quantum effects, and the universe can undergo eternal inflation \cite{Linde:1986fd, Boubekeur:2005zm, Barenboim:2016mmw}. This does not happen in the observationally relevant range of field values for Higgs inflation, but the field may have been closer to the hilltop earlier. 

In the Higgs case, a hilltop can form at small, intermediate or large field values. Let us go through these regions in turn.

\section{Small and intermediate field values} \label{sec:small_and_interm}

\subsection{Small field values} \label{sec:SM}

Let us first consider the possibility of hilltop inflation in the regime $h\ll1/\xi$, where the non-minimal coupling is irrelevant and only the usual SM loop corrections come into play. It is well known that the Higgs potential may have a secondary local minimum due to quantum corrections \cite{Degrassi:2012ry, Buttazzo:2013uya, Espinosa:2015a, Espinosa:2015b, Iacobellis:2016, Butenschoen:2016lpz, Espinosa:2016}. If so, there is a hilltop between the electroweak and the secondary minimum. For successful hilltop inflation, the potential must be flat enough near the hilltop so that not only $\epsilon$ but also the other slow-roll parameters are small, and the normalization of the perturbations \eqref{eq:pert_norm} must be correct at the pivot scale corresponding to the right number of e-folds \re{Npivot}.

To see if these criteria can be satisfied, we scan over the values of $m_H$ and $m_t$ as explained under equation \eqref{eq:SM_bestfit_vals}, and construct the corresponding SM Higgs potentials using the approximation presented in section \ref{sec:quantumcorrs}. As the first criterion after the existence of the hilltop, we demand that the slow-roll parameter $\eta$ should be small there. It turns out that this only happens very close to a critical line on the $(m_H,m_t)$ plane, as shown in figure \ref{fig:crit_line}. On the critical line the hilltop is transformed into a saddle point, i.e. the second derivative of the potential also vanishes.

We then scan over the critical line to see if the rest of the criteria are satisfied anywhere. It is  easy to find the point of correct normalization \eqref{eq:pert_norm} on the potential and calculate the slow-roll parameters and observables there. We find no viable regions: for example, we always have $\alpha_s < -90$. For SM without non-minimal coupling, in the approximation used here, hilltop inflation is not possible.

SM Higgs inflation without non-minimal coupling has been studied before at an inflection point, where the second derivative of the potential vanishes \cite{Hamada:2013mya,Fairbairn:2014nxa}, including the case of a saddle point \cite{Hamada:2013mya}. The results of \cite{Hamada:2013mya} were based on approximating the Higgs potential near the saddle point, whereas we have constructed the entire potential starting from the electroweak scale masses and couplings. Nevertheless, the conclusion is the same: we find no inflationary scenario that agrees with observations.

\begin{figure}
\begin{center}
\includegraphics[scale=0.7]{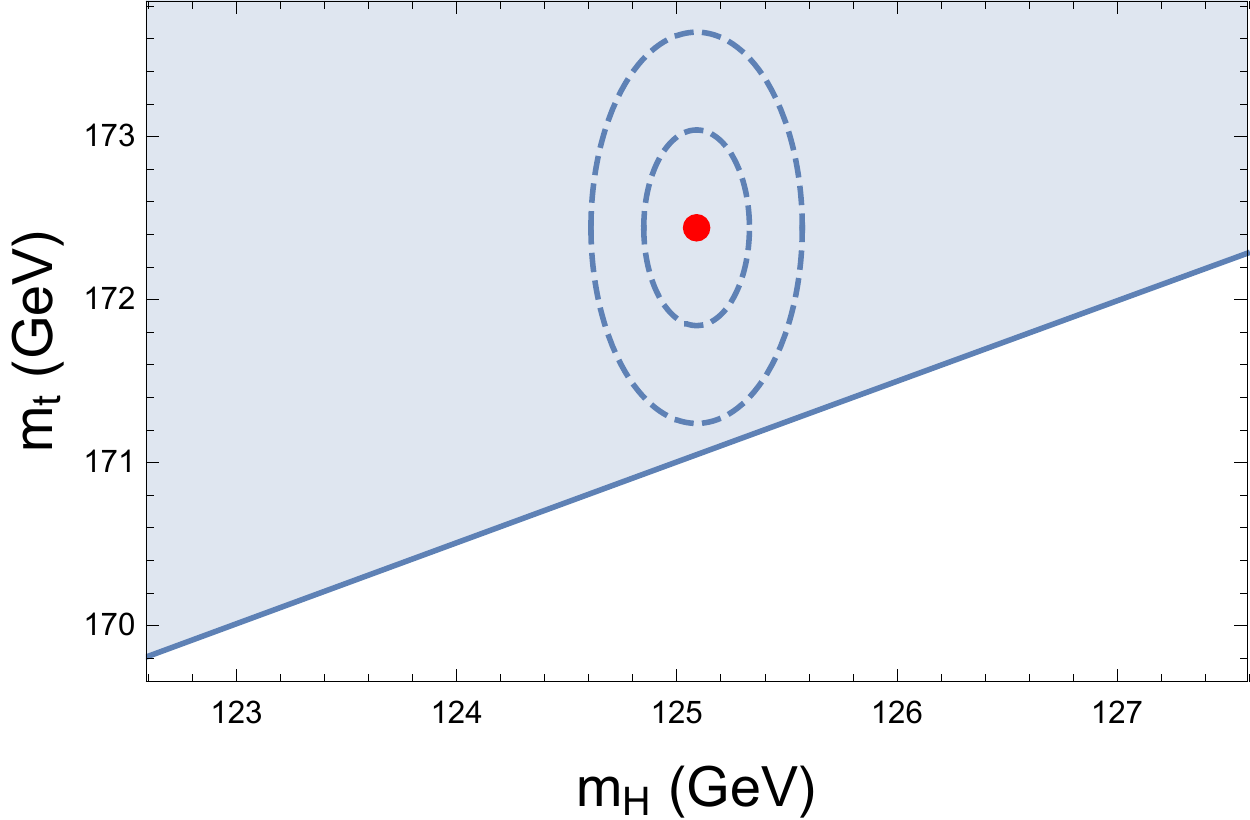}
\end{center}
\caption{In the SM, there is a hilltop for values of the Higgs mass $m_H$ and top mass $m_t$ in the shaded region. On the critical line between the regions, there is a saddle point. The red dot corresponds to the mean values \eqref{eq:SM_bestfit_vals}, and the dashed lines show the 1$\sigma$ and 2$\sigma$ contours, where the 1$\sigma$ uncertainty of $m_H$ is given by \eqref{eq:SM_bestfit_vals}, and the 1$\sigma$ uncertainty of $m_t$ is 0.6 GeV (see the discussion below \eqref{eq:SM_bestfit_vals}). The distance from the mean values to the critical line is between 2$\sigma$ and 3$\sigma$.
}
\label{fig:crit_line}
\end{figure}

\subsection{Intermediate field values} \label{sec:intermediate}

Let us now consider hilltop inflation in the regime $1/\xi \lesssim h \lesssim 1/\sqrt{\xi}$ by approximating the potential there by \eqref{eq:interm_pot}, with a continuous step in the coupling $\lambda$ given by \eqref{eq:lambda_jump}. The tree-level potential with constant $\lambda$ grows with $\chi$, but if $\lambda$ decreases rapidly enough due to the step, the potential can develop a local maximum.

However, as the step \eqref{eq:lambda_jump} is not very steep, we need $\Delta\lambda$ to be almost as big as $\lambda_0$ in \eqref{eq:lambda_jump} to get a maximum. In addition, $\Delta\lambda$ has to be finely tuned to make the slow-roll parameter $\eta$ small at the hilltop. The only agreeable potential is again of the saddle point type, and the field value $h$ there is typically of the order $1/\xi$ or smaller. The good news is that for any given $\xi$, successful inflation is always possible when the other parameters are tuned finely enough. The reason is that the shape of the potential is determined only by the ratio $\lambda_0/\Delta\lambda$, and the height can be adjusted separately by changing $\lambda_0$.

The parameters can be tuned as follows. First, pick a value for $\xi$. Then, adjust the ratio $\lambda_0/\Delta\lambda$ to get a saddle point. Next, find the point on the potential curve that corresponds to the number of e-folds \re{Npivot}. In slow-roll, this condition is independent from the normalization of the potential, as can be seen from equation \eqref{eq:efolds}. Finally, fix $\lambda_0$ so that the amplitude of the perturbations matches observations.

When $\xi$ is large, the modes in the observable region are generated very close to the saddle point, and the potential can be approximated as
\begin{gather}
	\label{eq:potential_between_approx}
	U(\chi)=U_0-\frac{1}{6}U'''_0(\Delta\chi)^3 \ ,
\end{gather}
where the subscript zero henceforth denotes value at the hilltop, $\Delta\chi \equiv \chi_0-\chi$ and $U'''_0$ is the third derivative of the potential at the hilltop. The constants $U_0$ and $U'''_0$ cannot be easily approximated from \eqref{eq:interm_pot}, but we determine them numerically. Using the slow-roll equations \re{eq:efolds}, \eqref{eq:sr_parameters}, we get the inflationary observables in terms of the number of e-folds:
\begin{equation} \label{eq:btw_observables}
\begin{alignedat}{2}
	n_s = 1-\frac{4}{N} \ , \qquad  r = \frac{8 (\Delta\chi)^2}{N^2} \ , \qquad
	\alpha_s = -\frac{4}{N^2} \ , \qquad \beta_s = -\frac{8}{N^3} \ ,
\end{alignedat}
\end{equation}
where at the pivot scale $\Delta\chi = \frac{2U_0}{N U'''_0}$. The bad news is that the analytical estimate \re{eq:btw_observables} for the spectral index, $n_s\approx0.92$, is far outside the observational bounds. This is true both in the metric and the Palatini case, as the approximation \eqref{eq:potential_between_approx} works equally for both. Therefore, and because our approximation for the shape of the potential in the intermediate region is in any case not robust, we do not perform a numerical scan over the parameters. 

It cannot be ruled out that quantum corrections could lead to a shape of the potential that would support successful hilltop inflation in this region, which has turned to be particularly interesting for inflation at a near-inflection point \cite{Allison:2013uaa, Bezrukov:2014bra, Hamada:2014iga, Bezrukov:2014ipa, Rubio:2015zia, Fumagalli:2016lls, Enckell:2016xse, Rasanen:2017ivk}. However, the effect of the UV completion of the theory on renormalization group running would need to be carefully controlled. The fact that the power counting estimate of the scale where perturbative unitarity is lost is parametrically higher in the Palatini case may have an impact on the details of Higgs inflation in the intermediate regime \cite{Bauer:2010jg}.

\section{Large field values} \label{sec:largefield}

\subsection{Analytical approximation} \label{sec:large_field_anal}

\subsubsection{Metric formulation} \label{sec:metric_anal}

Let us now look at inflation in the regime $h\gg1/\sqrt{\xi}$, which is also the setting for Higgs tree-level plateau inflation \cite{Bezrukov:2007ep}. Now quantum corrections produce a hilltop there, modifying the predictions. We first consider the metric formulation and then discuss what changes in the Palatini case.

To gain analytical understanding, we first consider just the tree-level potential augmented with a running $\lambda$ that produces a hilltop. We will apply the full machinery of section \ref{sec:quantumcorrs} in the next section.
It is useful to define the new parameter \cite{Fumagalli:2016lls}
\begin{equation} \label{eq:delta}
	\delta \equiv \frac{1}{\xi h^2} \ ,
\end{equation}
which is small in the region $h \gg 1/\sqrt{\xi}$. We calculate analytically to lowest order in $\delta$.

The potential is
\begin{equation} \label{eq:large_appr_pot}
	U=\frac{\lambda(\chi)}{4}F^4 \ , \qquad F=\frac{1}{\sqrt{\xi\qty(1+\delta)}} \ ,
\end{equation}
with $\lambda$ running as
\begin{equation} \label{eq:large_lambda_running}
	\frac{d\lambda}{d\chi} = \frac{d\lambda}{d\ln \mu}\frac{d\ln \mu}{d\chi} = \beta_\lambda \frac{F'}{F} \ .
\end{equation}
The derivative of the potential with respect to $\chi$ is thus
\begin{equation} \label{eq:large_appr_pot_deriv}
	U'=\frac{F^3F'}{4}\qty(4\lambda + \beta_\lambda) \ .
\end{equation}
This vanishes at the hilltop, so we can approximate
\begin{equation} \label{eq:large_lambda_appr}
	\lambda = \lambda_0 - 4\lambda_0\ln\frac{\mu}{\mu_0} \ ,
\end{equation}
where the renormalization scale \eqref{eq:renormscale_chiral} in terms of $\delta$ \eqref{eq:delta} is
\begin{equation} \label{eq:large_mu_lambda}
	\mu = \frac{\gamma}{\sqrt{\xi\qty(1+\delta)}} \ .
\end{equation}
During inflation, $\delta$ is between zero and one, and we have
\begin{equation} \label{eq:large_delta_mu}
	\Delta \ln \mu \ll \ln{\frac{\gamma}{\sqrt{\xi(1+0)}}} - \ln{\frac{\gamma}{\sqrt{\xi(1+1)}}} = \frac{1}{2}\ln{2} \approx 0.35 \ .
\end{equation}
This is a small number, so truncating the Taylor series \eqref{eq:large_lambda_appr} to linear order seems valid.

The free parameters that determine the shape of the potential are now $\delta_0$, $\lambda_0$ and $\xi$. Out of these, $\lambda_0$ cancels out when calculating the slow-roll parameters in our current approximation.
By assuming $\xi \gg 1$ we get for the slow-roll parameters at scale $\delta > \delta_0$, to lowest order in $\delta$ and $\delta_0$:
\begin{equation} \label{eq:large_sr_parameters}
\begin{alignedat}{2}
	\epsilon &= \frac{16}{3}\delta^2(\delta-\delta_0)^2 \ , \qquad &\eta &= -\frac{8}{3}\delta(2\delta - \delta_0) \ , \\
	\zeta &= \frac{64}{9}\delta^2(\delta - \delta_0)(4\delta - \delta_0) \ , \qquad	&\varpi &= -\frac{512}{27}\delta^3(\delta-\delta_0)^2(8\delta-\delta_0) \ .
\end{alignedat}
\end{equation}
The number of e-folds \eqref{eq:efolds} corresponding to $\delta$ is
\begin{equation} \label{eq:NcSM}
	N = \frac{3}{8\delta_0^2}\left( \ln\frac{\delta}{\delta-\delta_0} - \frac{\delta_0}{\delta}\right) \ .
\end{equation}
Fixing $N$ and $\delta_0$ fixes $\delta$ and thus the slow-roll parameters and the observables \eqref{eq:sr_parameters}. Figure \ref{fig:large_analytic_pics} illustrates the behaviour of $n_s$ and $r$.

\begin{figure}
\begin{center}
\includegraphics[scale=0.7]{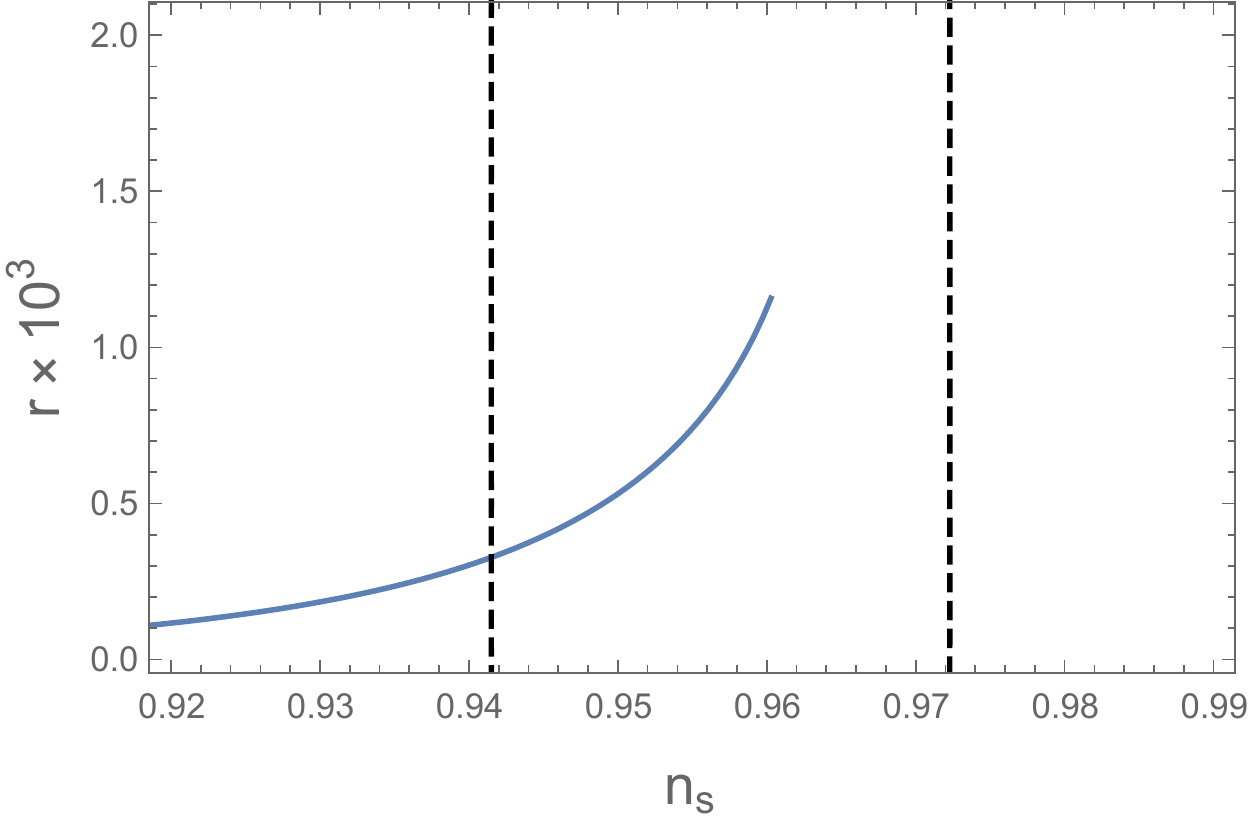}
\end{center}
\caption{The spectral index $n_s$ and tensor-to-scalar ratio $r$ in the analytical approximation of section \ref{sec:large_field_anal}, for the number of e-folds \eqref{Npivot}. The dashed lines indicate the observational 2$\sigma$ range of $n_s$ given by \eqref{eq:sr_obs_n}. Motion along the line corresponds to changing the value of $\delta_0$, the endpoint at $n_s = 0.96$ corresponding to the infinite field limit $\delta_0\rightarrow0$.}
\label{fig:large_analytic_pics}
\end{figure}

For $n_s$ and $r$, the large field limit $\delta_0 \to 0$ is of particular interest:
\begin{equation} \label{eq:limit_d0_to_0}
	n_s \xrightarrow{\delta_0 \to 0} 1 - \frac{2}{N} \ , \qquad r \xrightarrow{\delta_0 \to 0} \frac{3}{N^2} \ .
\end{equation}
In this limit the hilltop moves to infinite field values, so we no longer have a local maximum. Instead the running of the coupling just makes the potential flatter than in tree-level Higgs inflation on the plateau. The result for $n_s$ approaches the tree-level case \eqref{eq:higgsinfl_standard_results}, while the limiting value of $r$ is a factor of four smaller. Spanning the interval between the tree-level and hilltop cases would require deforming the potential so that there is no hilltop; this region has been covered in \cite{Enckell:2016xse}. The results \eqref{eq:limit_d0_to_0} are the upper limits for $n_s$ and $r$.
The values of $\alpha_s$ and $\beta_s$ calculated from \eqref{eq:large_sr_parameters} are approximately constant for fixed $N$ in the allowed range of $n_s$. They are close to the tree-level results \re{eq:higgsinfl_standard_results_N50}.

Precise results for the number of e-folds \re{Npivot} give
\begin{equation} \label{eq:lergefield_anal_result_ns}
	n_s < 0.96 \ .
\end{equation}
Restricting $n_s$ to the observational 95\% confidence interval \re{eq:sr_obs_n}, we get
\begin{equation} \label{eq:largefield_anal_results_other}
\begin{gathered}
	3.3 \times 10^{-4} < r < 1.2 \times 10^{-3} \ , \\
	-0.83 \times 10^{-3} < \alpha_s < -0.79 \times 10^{-3} \ , \quad -3.4 \times 10^{-5} < \beta_s < -3.2 \times 10^{-5} \ .
\end{gathered}
\end{equation}
The observed perturbation amplitude \eqref{eq:pert_norm} gives a relation between $\lambda_0$, $\xi$ and $r$:
\begin{equation}
	\xi = \sqrt{\frac{4 \lambda_0}{5.2 \times 10^{-7} r}} \ .
\end{equation}
Restricting $r$ to the bounds of \eqref{eq:largefield_anal_results_other} and $\lambda_0$ to the interval from zero to one, we get
\begin{equation}
	0 < \xi <  1.5 \times 10^5 \ .
\end{equation}

\subsubsection{Palatini formulation} \label{sec:palatini_anal}

\begin{figure}
\begin{center}
\includegraphics[scale=0.45]{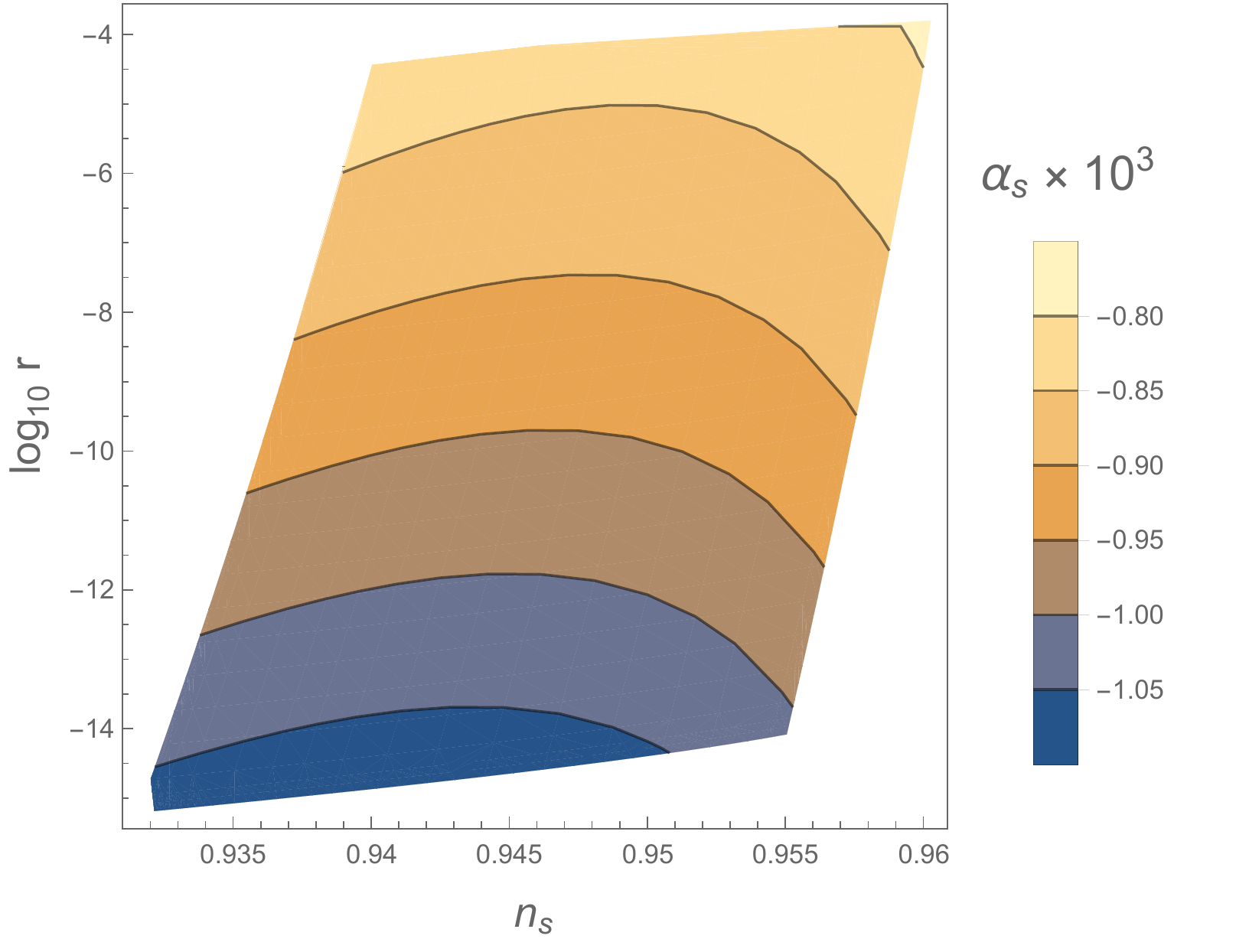}
\includegraphics[scale=0.45]{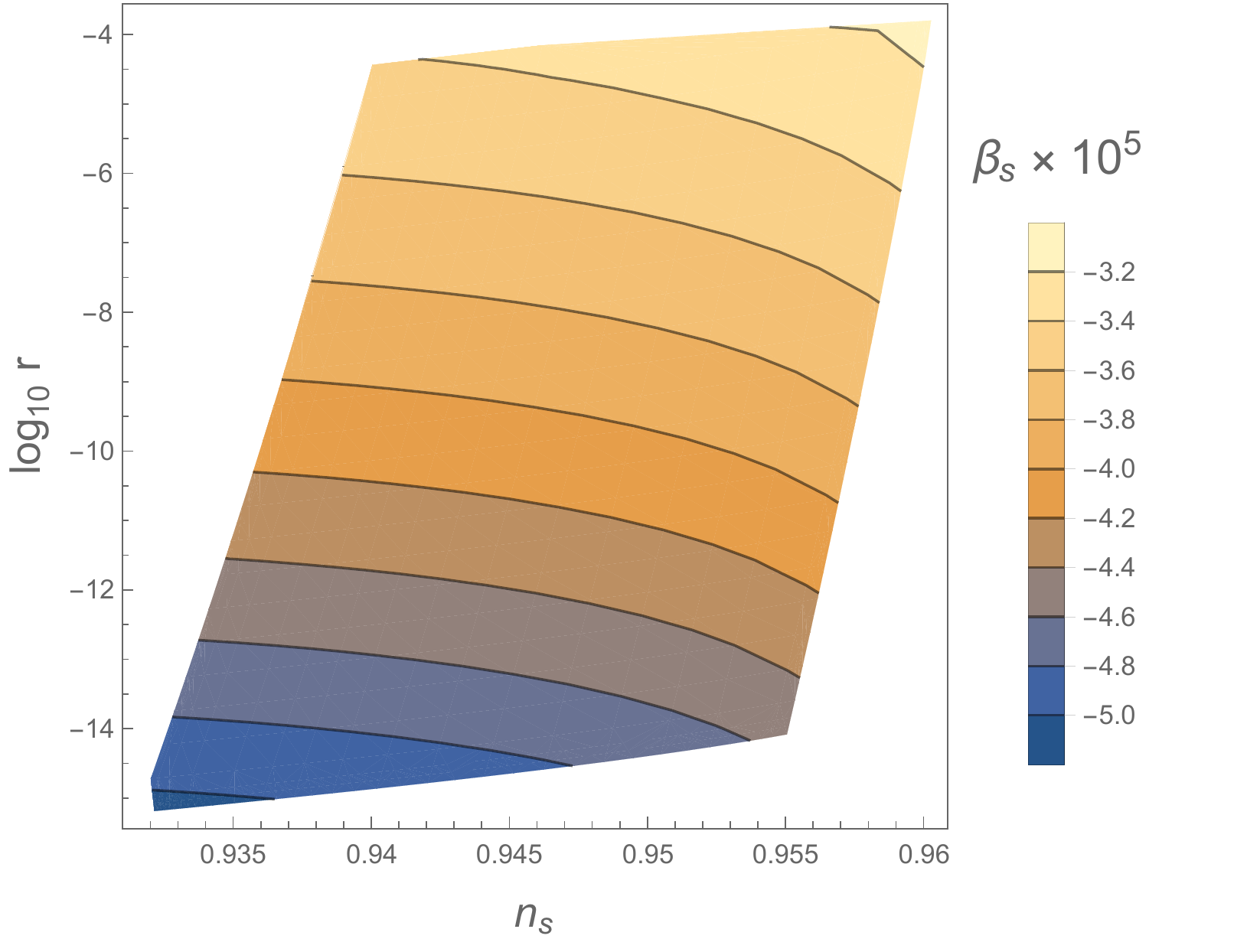}
\end{center}
\caption{Results of the analytical approximation for $\alpha_s$ and $\beta_s$ as a function of $n_s$ and $r$ in the Palatini case.}
\label{fig:palatini_anal_pics}
\end{figure}

In the large field region $h\gg1/\sqrt{\xi}$, the Palatini formulation differs from the metric formulation at leading order only in that all derivatives with respect to $\chi$ gain a factor of $\sqrt{6\xi}$ (see \eqref{eq:dchi_dh}), so the results \eqref{eq:large_sr_parameters}, \eqref{eq:NcSM} become
\begin{equation} \label{eq:large_sr_parameters_pala}
\begin{alignedat}{2}
	\epsilon &= 6\xi \times \frac{16}{3}\delta^2(\delta-\delta_0)^2 \ , \quad
	&\eta &= 6\xi \times \qty(-\frac{8}{3})\delta(2\delta - \delta_0) \ , \\
	\zeta &= (6\xi)^2 \times \frac{64}{9}\delta^2(\delta - \delta_0)(4\delta - \delta_0) \ , \quad
	&\varpi &= (6\xi)^3 \times \qty(-\frac{512}{27})\delta^3(\delta-\delta_0)^2(8\delta-\delta_0) \ , \\
	N &= \frac{1}{6\xi} \times \frac{3}{8\delta_0^2}\left( \ln\frac{\delta}{\delta-\delta_0} - \frac{\delta_0}{\delta}\right) \ .
\end{alignedat}
\end{equation}
There are now two free parameters that characterize the dynamics, $\delta_0$ and $\xi$, whereas in the metric case $\xi$ played no role. Thus in the analogue of figure \ref{fig:large_analytic_pics}, the single line expands to a two-dimensional area. However, in both cases, the observables $\eta$, $\zeta$, $\varpi$, and $N$ have the same expression in terms of the variables $\delta_0/\delta$ and $(1+p[\xi-1])\delta_0^2$ (recall that $p$ is 0 in the metric formulation and 1 in the Palatini formulation), so $\alpha_s$ and $\beta_s$ remain, to a first approximation, unchanged from the metric case. However, $r$ is different: it is suppressed by $6\xi$, as in the tree-level plateau case \cite{Bauer:2008zj}.

The precise results are shown in figure \ref{fig:palatini_anal_pics}. The allowed $(n_s,r)$ parameter space is bounded from above by the requirement $\xi>1$, from below by $\lambda_0<1$, from right by the limit $\delta_0\to0$, and from left by the choice to only consider a branch of the solutions close to the observed value $n_s\approx0.96$. Since numerical considerations add more such constraints, we don't here quote any numbers from the analytical approach, but instead refer to the results of the full numerical calculation of section \ref{sec:palatini_numer}.

\subsection{Numerical results} \label{sec:large_field_numer}

\subsubsection{Metric formulation} \label{sec:metric_numer}

For the numerical results, we abandon the approximations of the previous section, and consider the full loop-corrected effective potential with the parameters run with the renormalization group equations as described in section \ref{sec:quantumcorrs}. We use the input parameters of the numerical calculation to deduce the form of the potential $U(\chi)$ in the large field limit. Then we calculate the evolution of the field using the full time evolution equations \eqref{eq:friedmann_scalar_eq} with derivatives taken using \eqref{eq:dchi_dh}, starting from slow-roll near the hilltop up to the end of inflation. This gives the inflationary observables plotted in figures \ref{fig:large_numeric_pics_metric_1} and \ref{fig:large_numeric_pics_metric_2}.

The calculation takes as input parameters the values $\delta_0$ and $\lambda_0$ defined in the previous section and the central values of the low-energy SM parameter \eqref{eq:SM_bestfit_vals}. Note that there are now two free parameters at the inflationary scale, instead of just one as in the analytical case of section \ref{sec:large_field_anal}. This is because in section \ref{sec:large_field_anal}, the dependence on $\lambda_0$ cancelled out in the simple expressions, but this is no longer true when the quantum corrections and the detailed running are taken into account. Nevertheless, we can fix all of the free parameters needed for $U(\chi)$.

The non-minimal coupling $\xi$ is determined iteratively to achieve the observed normalization for perturbations \eqref{eq:pert_norm} at the pivot scale corresponding to the value of $N$ given in \re{Npivot}, with an initial guess given by the approximation of the previous section. As the initial conditions for the renormalization group equations, we take the measured mean values \re{eq:SM_bestfit_vals}, together with the values for other electroweak observables given in the code \cite{SMRunningCode}, as discussed in section \ref{sec:quantumcorrs}. We run the couplings with the SM renormalization group equations to the matching scale \re{eq:renormscale_chiral}, where we switch to the chiral SM running. The values of the couplings other than $\lambda$ and $y_t$ at the hilltop then depend only on the matching parameter $\gamma$ given in \re{eq:renormscale_chiral}. Requiring $U'(\chi_0)=0$ and $U_{\mathrm{1-loop}}(\chi_0)=0$ (as discussed below \eqref{eq:renormscale_chiral}), we have two equations that determine the two unknowns $\gamma$ and $y_{t0}$ and fix the potential. Afterwards, we run $\lambda$ and $y_t$ up from the electroweak scale and down from the hilltop to the matching scale \eqref{eq:jumpscale} separately, and compare them there to determine the jumps $\Delta\lambda$ and $\Delta y_t$.

For the mean values \re{eq:SM_bestfit_vals} of the Higgs and top mass, the Higgs potential becomes negative at $6\times10^{11}$ GeV. For $\xi\lesssim10^6$, this is below the matching scale \eqref{eq:jumpscale}, where the SM approximation should be valid. In the metric case, this happens in practically the whole parameter space, and in the Palatini case it happens in a sizeable part of it, as can be seen from figures \ref{fig:large_numeric_pics_metric_1} and \ref{fig:large_numeric_pics_palatini_1}. This would make the electroweak vacuum metastable, but as long as the lifetime is much longer than the age of the universe, there is no contradiction with late-time observations. Another possible problem is that after inflation the field may get stuck in the true vacuum instead of rolling to the electroweak vacuum. However, this depends on the details of the form of the potential, as the field may roll down the hill fast enough to pass over the true vacuum, or the vacuum may be uplifted by thermal corrections \cite{Bezrukov:2014ipa, Rubio:2015zia, Enckell:2016xse}. Nevertheless, to be sure that there is no problem, we could scan over the allowed range of Higgs and top masses under the condition that there is no extra minimum in the SM region. However, it turns out that the CMB observables are rather insensitive to these electroweak scale input parameters, as the jumps $\Delta\lambda$ and $\Delta y_t$ adjust to compensate for changes in them. In contrast, the presence of the extra vacuum is highly sensitive to the precise Higgs and top masses, and changing them within their 2$\sigma$ error regions eliminates the SM region extra vacuum. We therefore do not apply the extra condition that there is no extra vacuum in the SM region. We expect this to have negligible effect on the results. Below we give examples where the Higgs and top masses are consistent with the measured values, the jumps are zero, there is no extra vacuum in the SM region, and the CMB predictions agree with observations.

The crucial point is that we use the existence of the hilltop as an initial condition to fix the couplings $\lambda$ and $y_t$ at the hilltop, and only compare the results for $\lambda$ and $y_t$ with the SM afterwards. Higgs inflation at the hilltop has been studied before \cite{Fumagalli:2016lls, Rasanen:2017ivk}. However, in \cite{Fumagalli:2016lls} the hilltop was not investigated thoroughly, as much tuning and extensive searching through the parameter space is needed if the SM parameters are used as the starting point. We solve this problem by first constructing the hilltop and working backwards from there. In \cite{Rasanen:2017ivk} the potential was approximated with a simple form adapted to the minimum of $\lambda(h)$, which may not be valid at the hilltop.

In figure \ref{fig:large_numeric_pics_metric_1} we show the range of $n_s$, $r$, $\alpha_s$ and $\beta_s$ in terms of the parameters $\delta_0$ and $\xi$ in the metric case. For clarity, we have cut regions where $n_s<0.9$.
The value of $\xi$ is strongly correlated with $\lambda_0$ via the normalization of the amplitude. The value of $\xi$ is bounded from above by the fact that larger values would lead to a $\lambda_0$ that is so large that $y_t>1$ at the hilltop. The lower limit for $\xi$ comes from the condition that a hilltop won't be formed for values of $\lambda_0$ that are too small. On the right, we must have $\delta_0<1$ to be in the large field regime. On the left, the observables approach the analytical $\delta_0 \to 0$ limit, and there are no new features beyond the range plotted here. Hilltop inflation requires $180<\xi<1.7\times10^7$.

\begin{figure}
\begin{center}
\includegraphics[scale=0.4]{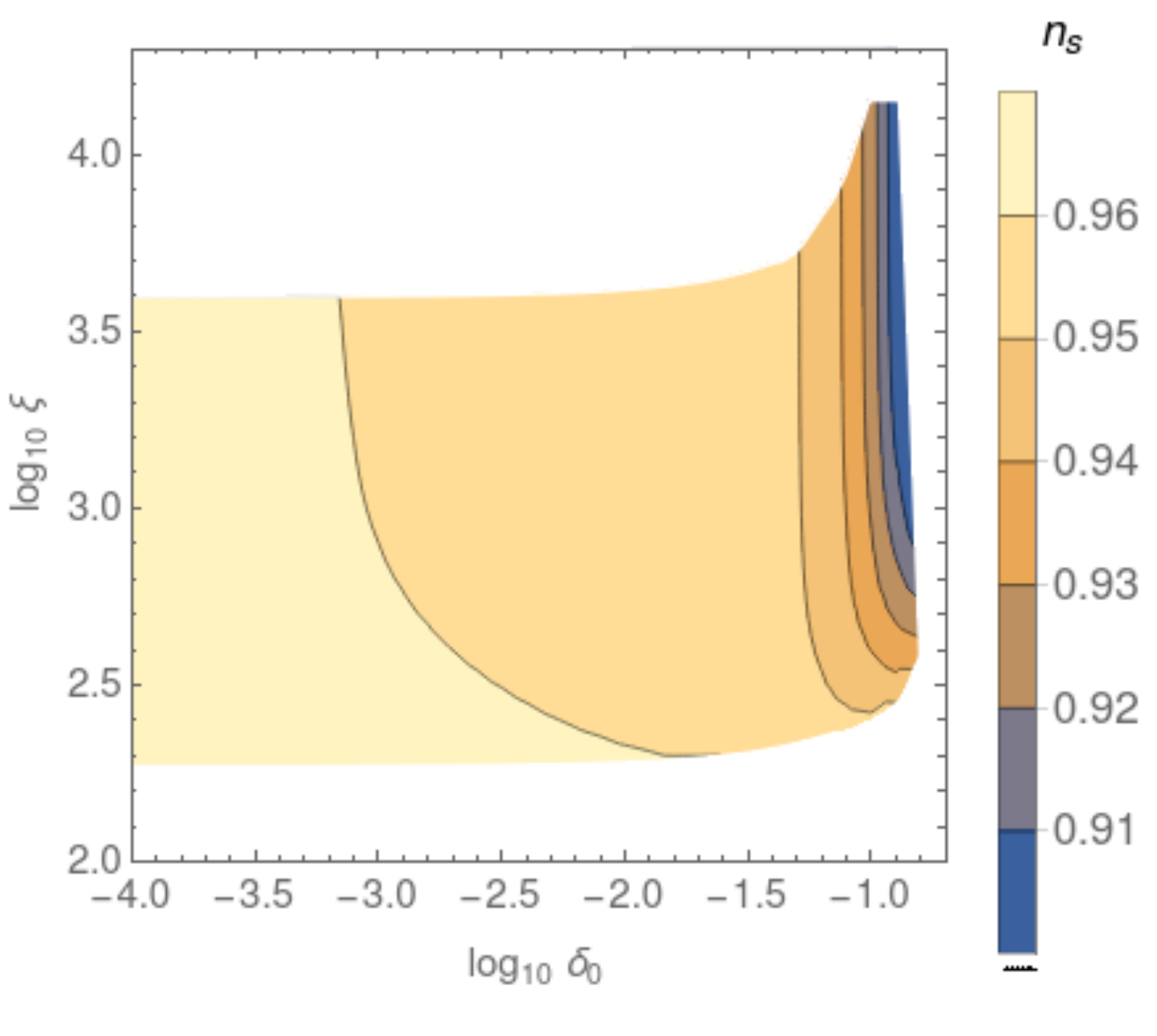}
\includegraphics[scale=0.4]{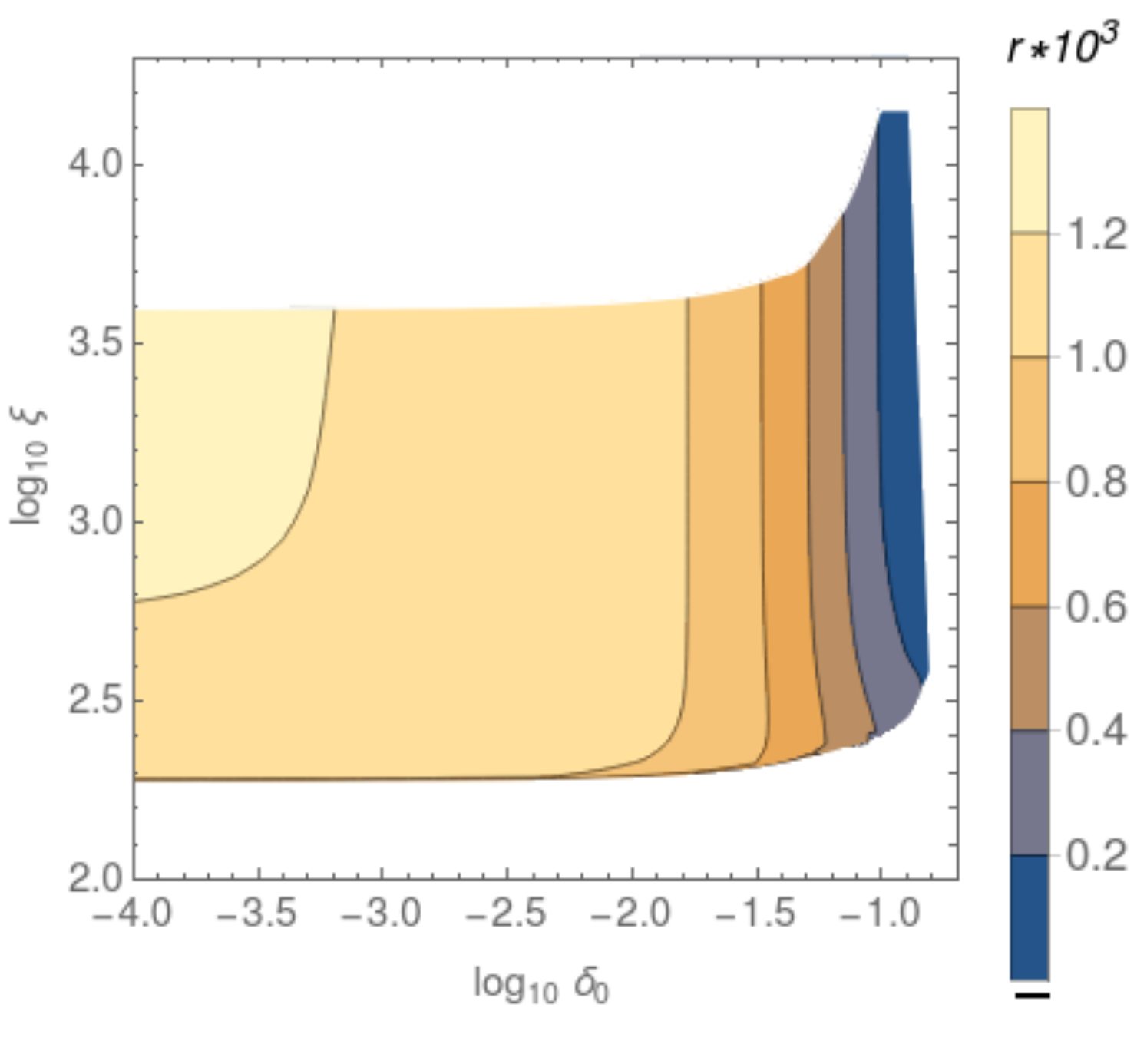}
\includegraphics[scale=0.4]{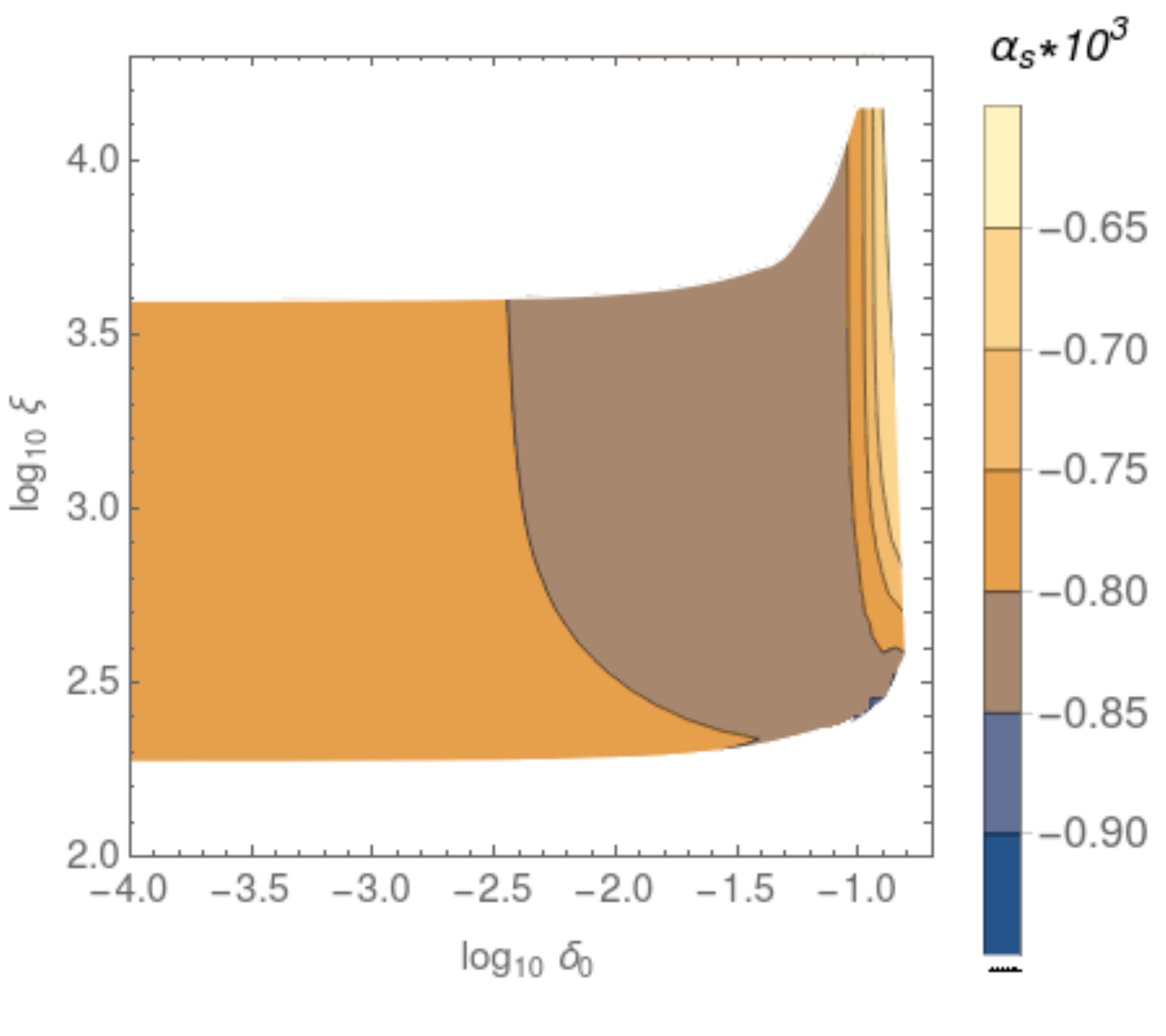}
\includegraphics[scale=0.4]{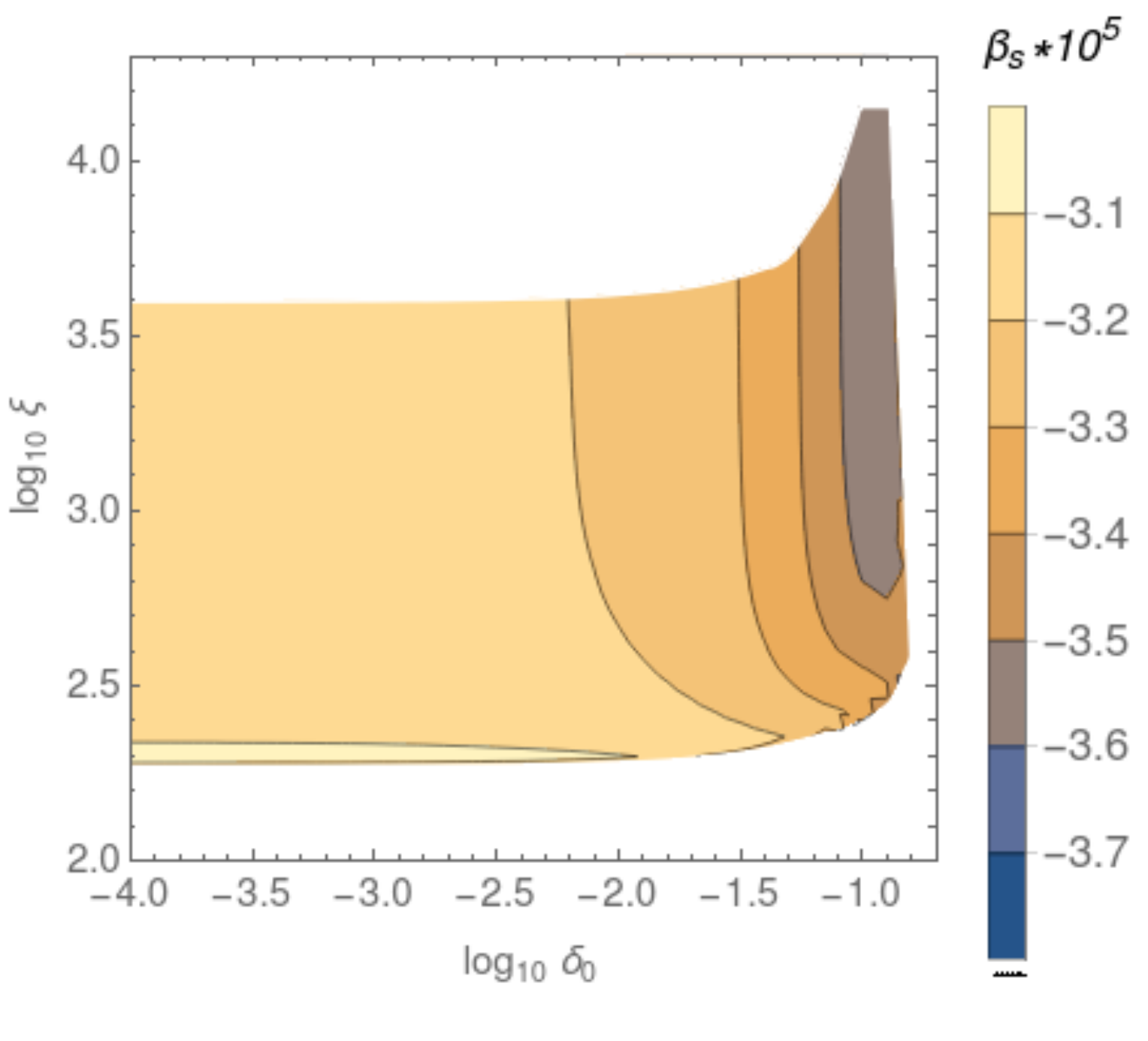}
\end{center}
\caption{Numerical values of $n_s$, $r$, $\alpha_s$ and $\beta_s$ as functions of $\delta_0$ and $\xi$ in the metric case. For clarity, we have cut regions where $n_s<0.9$.}
\label{fig:large_numeric_pics_metric_1}
\end{figure}

In figure \ref{fig:large_numeric_pics_metric_2} we show $\alpha_s$ and $\beta_s$ on the $(n_s,r)$ plane.  For the spectral index, we have $n_s\leq0.96$. Restricting $n_s$ to the observational 95\% confidence interval \re{eq:sr_obs_n}, the tensor-to-scalar ratio is $1.3\times10^{-4}<r<1.2\times10^{-3}$, always smaller than in the tree-level plateau case.
With the same restriction, the running and the running of the running are tightly determined, $-0.93\times10^{-3}<\alpha_s<-0.76\times10^{-3}$, $-3.8\times10^{-5}<\beta_s<-3.1\times10^{-5}$. They are close to the tree-level plateau results and well below the observational limits \re{eq:sr_obs_n}.

\begin{figure}
\begin{center}
\includegraphics[scale=0.3]{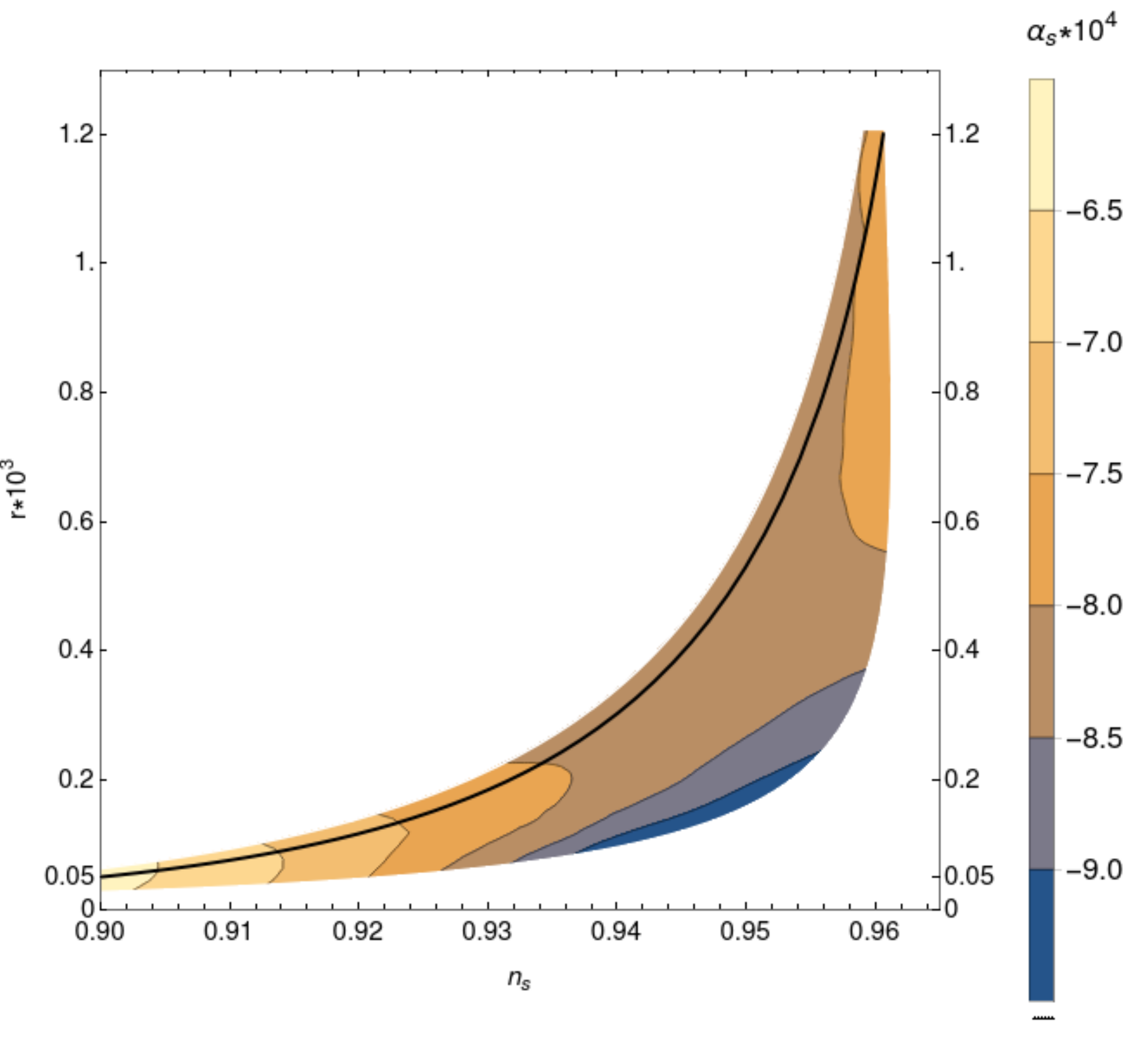}
\includegraphics[scale=0.3]{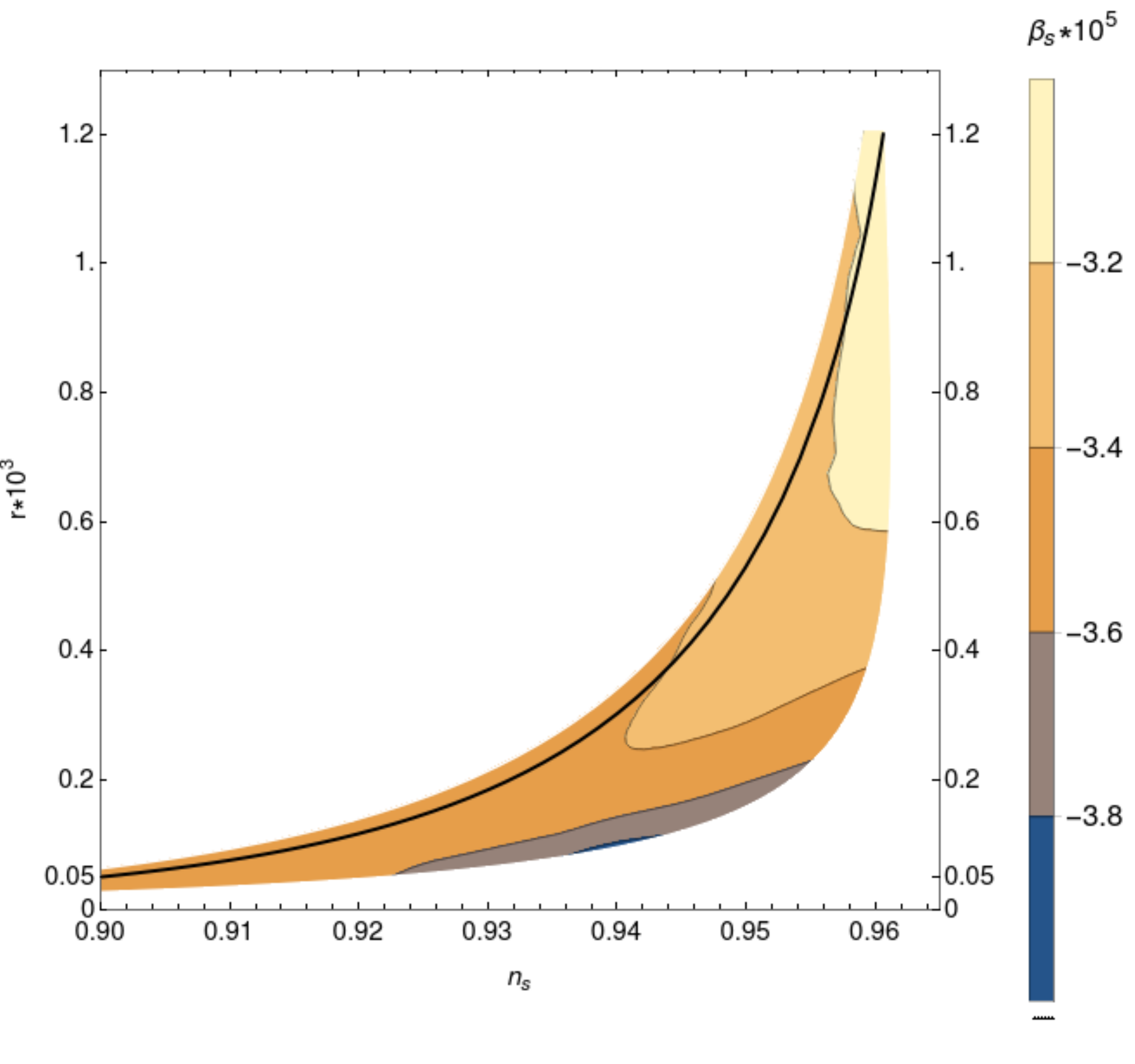}
\end{center}
\caption{Numerical values of $\alpha_s$ and $\beta_s$ on the $(n_s,r)$ plane in the metric case. The black line shows the analytical approximation plotted in figure \ref{fig:large_analytic_pics}.}
\label{fig:large_numeric_pics_metric_2}
\end{figure}

These results are consistent with the analytical approximation of section \ref{sec:metric_anal}. Points on the $(n_s,r)$ plane in figure \ref{fig:large_numeric_pics_metric_2} surround the analytical line of figure \ref{fig:large_analytic_pics}. When $n_s$ is restricted to the observational 95$\%$ range, the results for $r$, $\alpha_s$ and $\beta_s$ are almost identical to the analytical case \eqref{eq:largefield_anal_results_other}.

In \cite{Fumagalli:2016lls}, three data points were given for the hilltop case (in their table 1). There increasing $n_s$ increases $r$, and lowers $\xi$, which is consistent with our results. Our values for $n_s$ agree, but their corresponding values for $r$ are just below our results shown in figure \ref{fig:large_numeric_pics_metric_2}.

In \cite{Rasanen:2017ivk}, Higgs inflation at hilltop was considered using an analytical approximation for the potential, whereas we have used the full renormalization group running of the couplings and a loop-corrected potential. They found correlations between $n_s$, $r$ and $\xi$ that are similar to our results and to those of \cite{Fumagalli:2016lls}. The numerical values obtained for $n_s$ and $\alpha_s$ are also consistent with our results, but our lower range of $r$ is tighter by one order of magnitude, and our result for the absolute value of $\beta_s$ is an order of magnitude smaller.

These results were obtained using the central values of the low-energy SM parameters \eqref{eq:SM_bestfit_vals}, but they are robust to changes in $m_H$ and $m_t$. When $m_H$ and $m_t$ are varied within the 2$\sigma$ confidence intervals, the limits for the CMB observables change only in the second or higher digit, as the required jumps $\Delta\lambda$ and $\Delta y_t$ adjust to compensate. The average values of the jumps are of the order $0.02$ for $\Delta\lambda$ and $0.06$ for $\Delta y_t$.

It is interesting to ask whether the case with no jumps is observationally viable. As our method of setting initial conditions at the hilltop does not allow us to easily scan the electroweak observables, we have not done a comprehensive search of the parameter space. However, with the jumps put to zero, the values $m_H=125.53$ GeV, $m_t=171.29$ GeV (both inside the observational 95\% confidence intervals \eqref{eq:SM_bestfit_vals}) for example give $n_s=0.96$, $r=1.1\times10^{-3}$, $\alpha_s=-0.81\times10^{-3}$ and $\beta_s=-3.2\times10^{-5}$, all well within the observational bounds.

\subsubsection{Palatini formulation} \label{sec:palatini_numer}

In figure \ref{fig:large_numeric_pics_palatini_1} we show the range of $n_s$, $r$, $\alpha_s$ and $\beta_s$ in terms of the parameters $\delta_0$ and $\xi$ in the Palatini case.  The limits of the contours follow from the same constraints as in the metric case, but the value of $\delta_0$ is now bounded more tightly from above by the branch choice mentioned in section \ref{sec:palatini_anal}, which cuts out $n_s$ values smaller than about 0.94. As with inflation on the plateau or at the critical point \cite{Bauer:2008zj, Rasanen:2017ivk}, the Palatini formulation requires larger values of the non-minimal coupling, $1.0\times10^5<\xi<5.2\times10^8$, and $r$ is correspondingly highly suppressed.

\begin{figure}[!t]
\begin{center}
\includegraphics[scale=0.4]{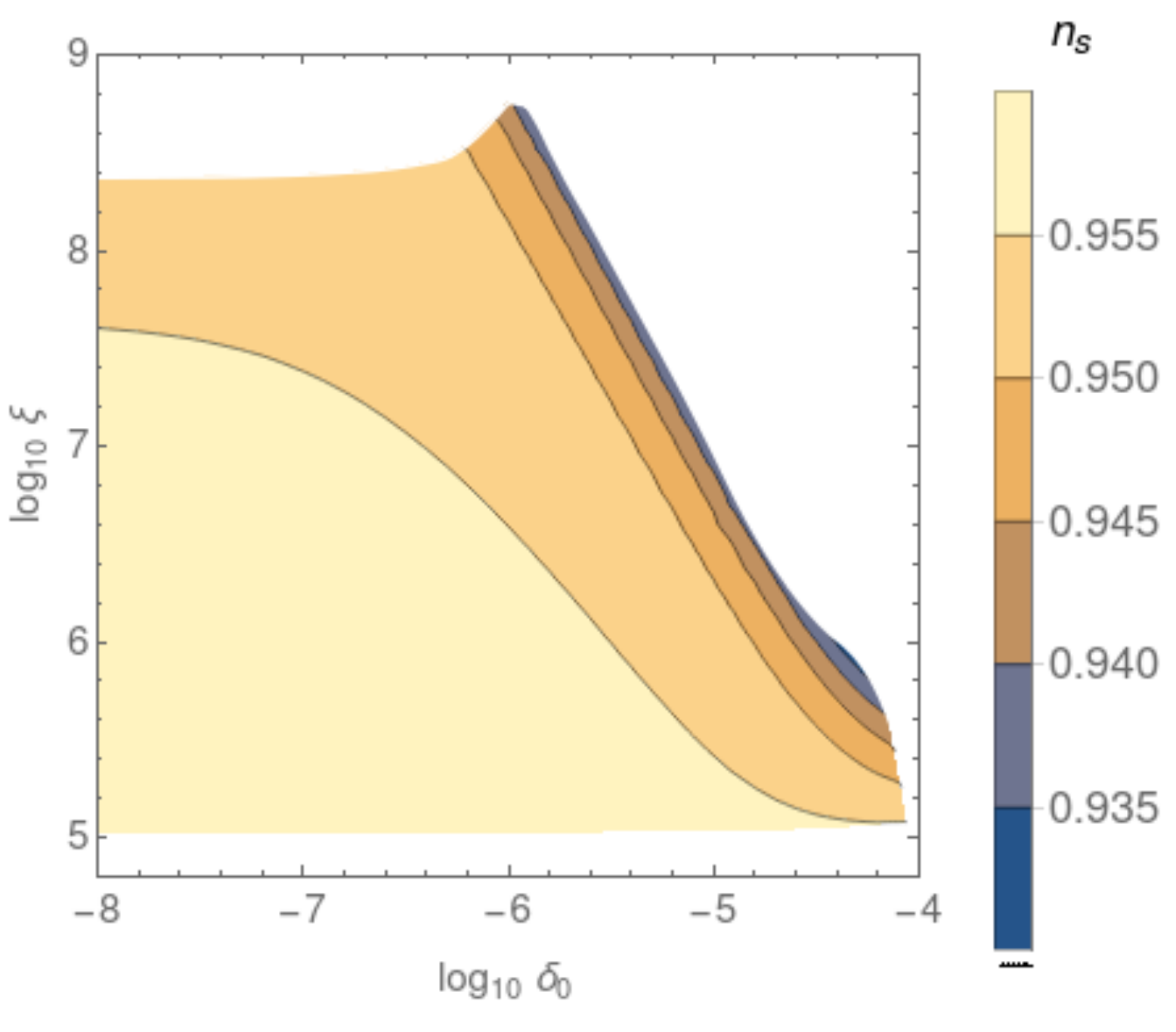}
\includegraphics[scale=0.4]{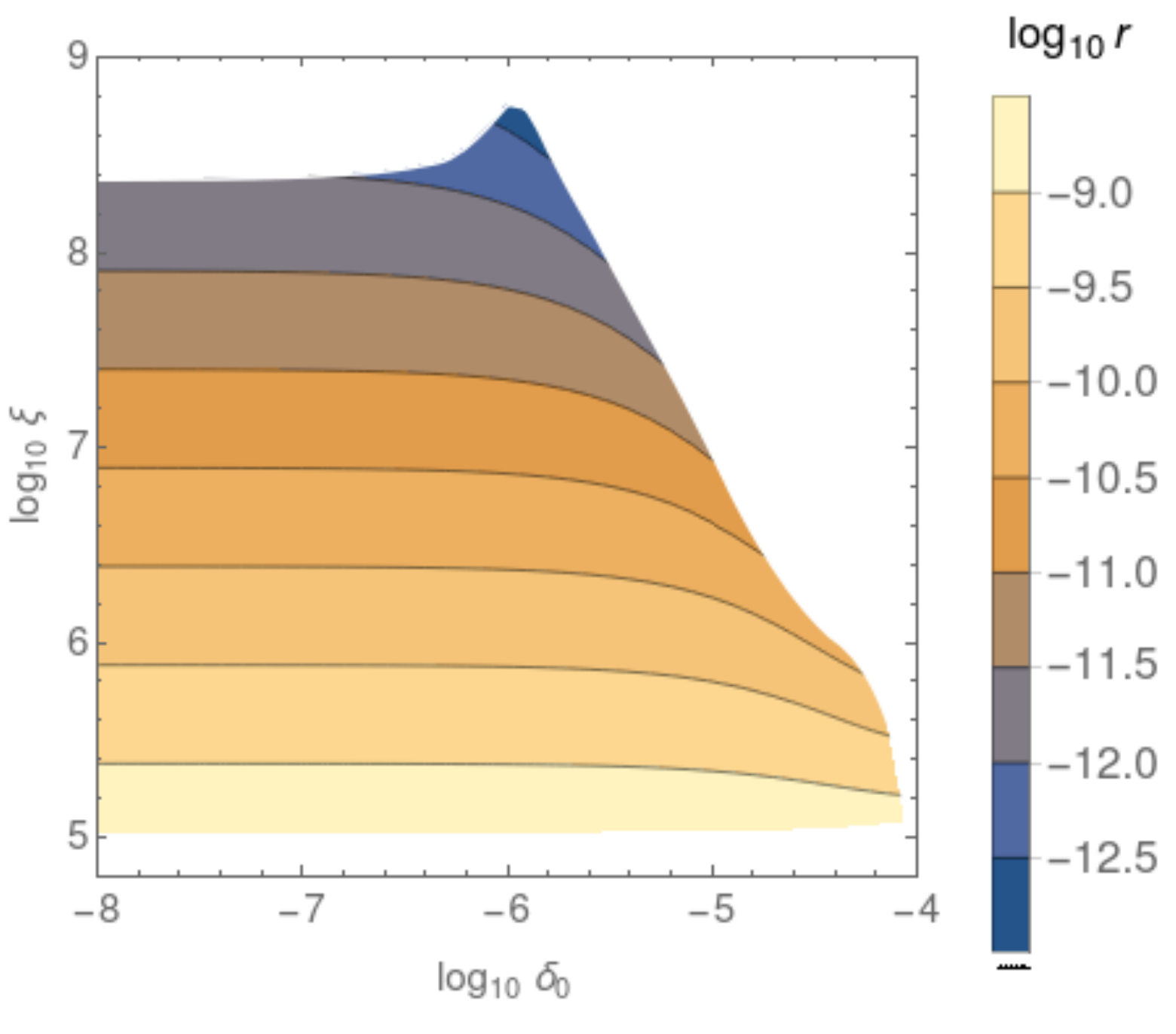}
\includegraphics[scale=0.4]{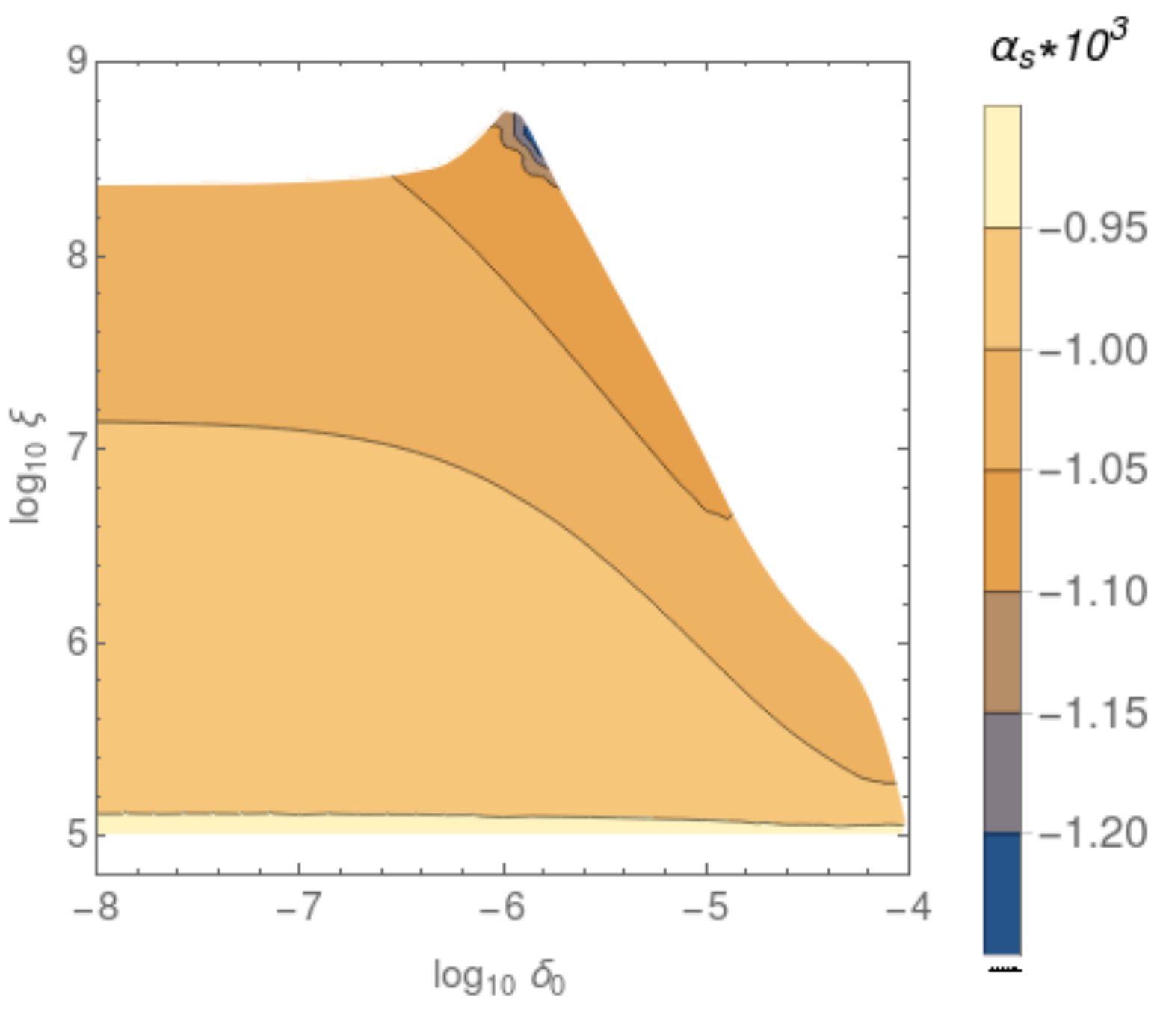}
\includegraphics[scale=0.4]{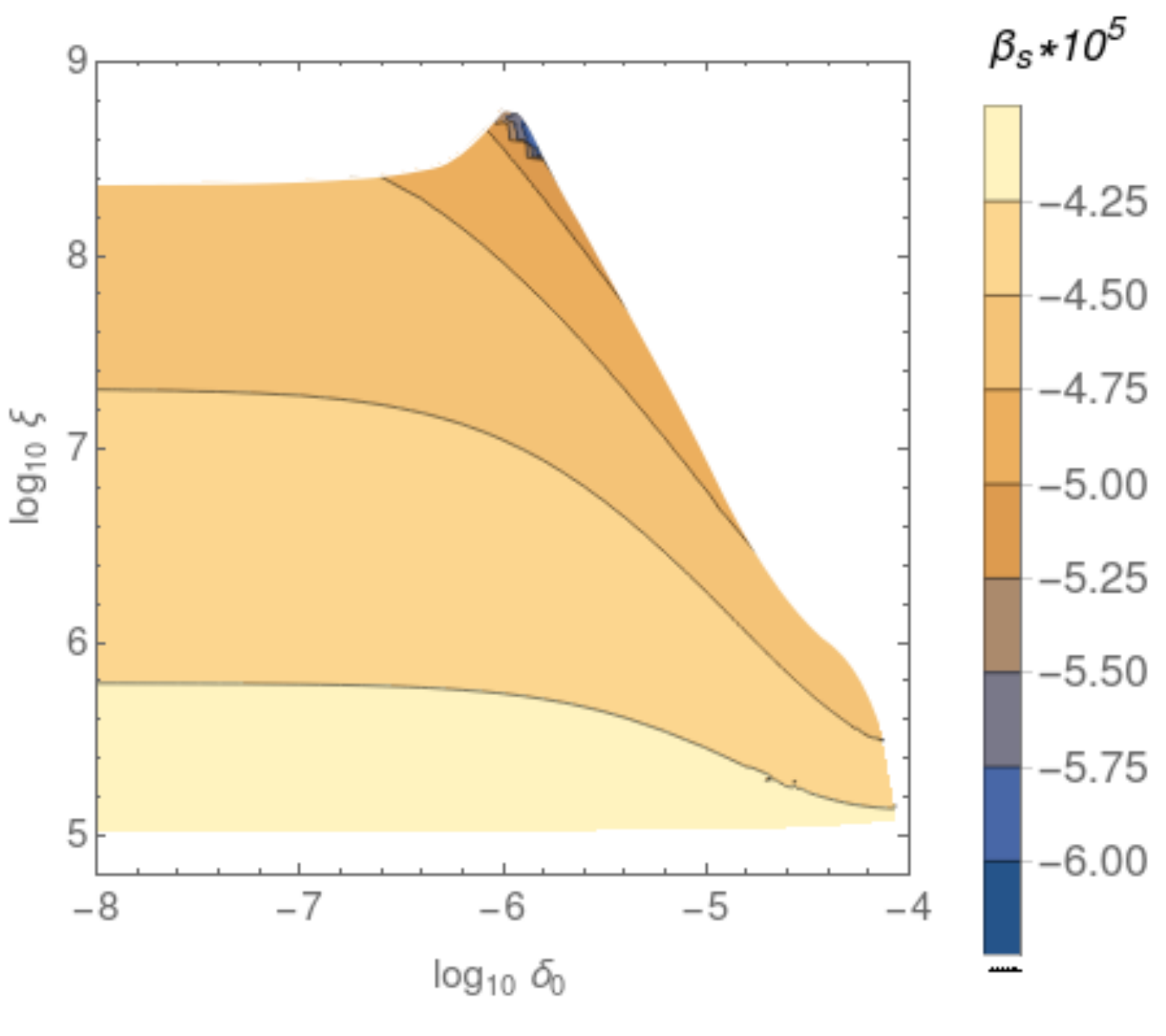}
\end{center}
\caption{Numerical values of $n_s$,  $r$, $\alpha_s$ and $\beta_s$ as functions of $\delta_0$ and $\xi$ in the Palatini case.}
\label{fig:large_numeric_pics_palatini_1}
\end{figure}

In figure \ref{fig:large_numeric_pics_palatini_2} we show $\alpha_s$ and $\beta_s$ on the $(n_s,r)$ plane. We have the same limit $n_s\leq0.96$ as in the metric case. Restricting $n_s$ to the observational 95\% confidence interval \re{eq:sr_obs_n}, the tensor-to-scalar ratio is $2.2\times10^{-13}<r<2.2\times10^{-9}$. Unlike in the metric formulation, this does not extend the range of predictions of the model from the plateau and critical point cases. Because of the strong dependence of $r$ on $\xi$, even the tree-level plateau inflation scans a wide range of possible values for $r$ \cite{Bauer:2008zj, Rasanen:2017ivk}. As in the metric case, the running and the running of the running conditioned on the observed range of $n_s$ are tightly determined, $-1.2\times10^{-3}<\alpha_s<-0.94\times10^{-3}$, $-5.5\times10^{-5}<\beta_s<-4.1\times10^{-5}$. They are of the same order of magnitude as in the metric case.

\begin{figure}[!t]
\begin{center}
\includegraphics[scale=0.3]{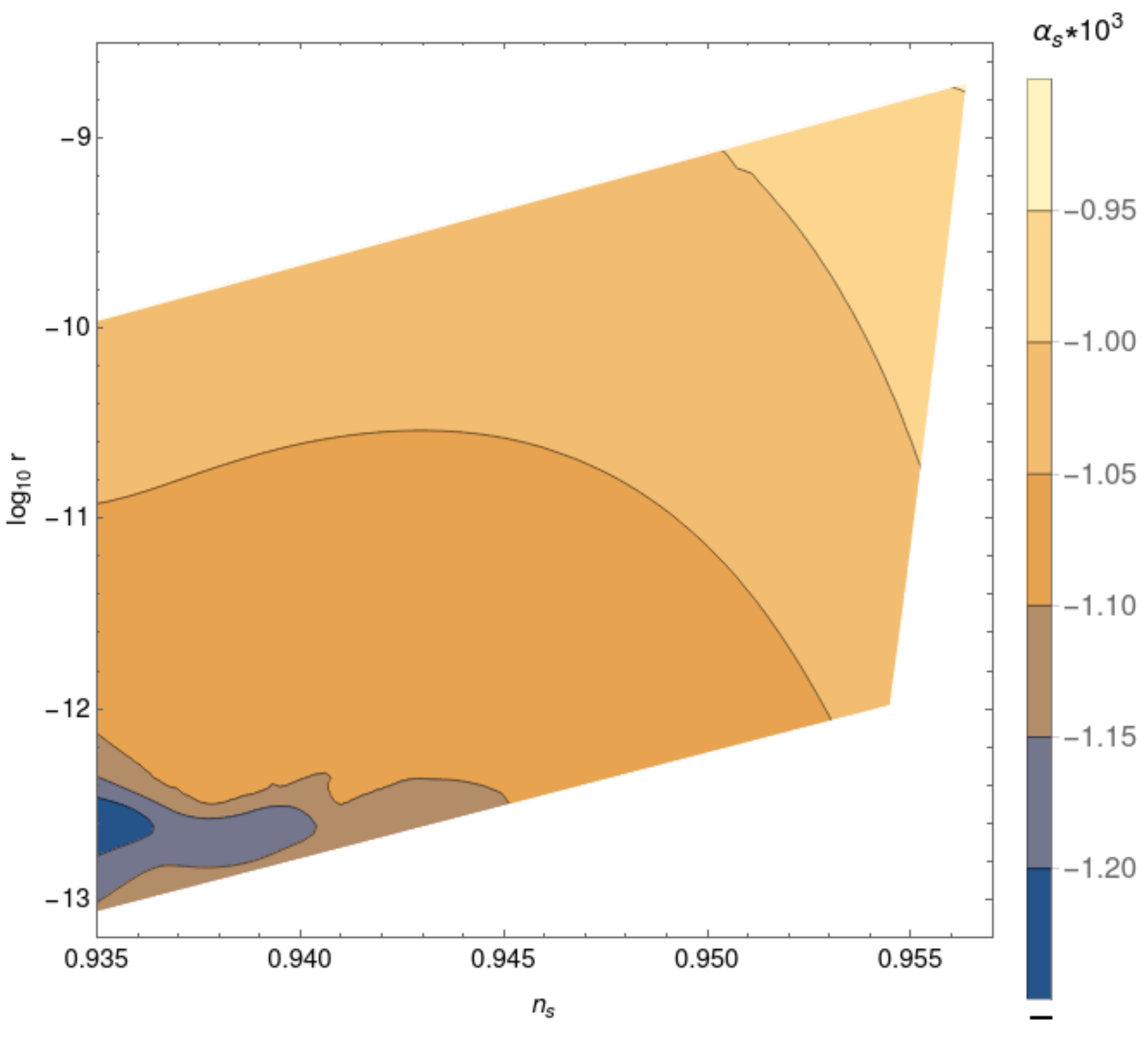}
\includegraphics[scale=0.3]{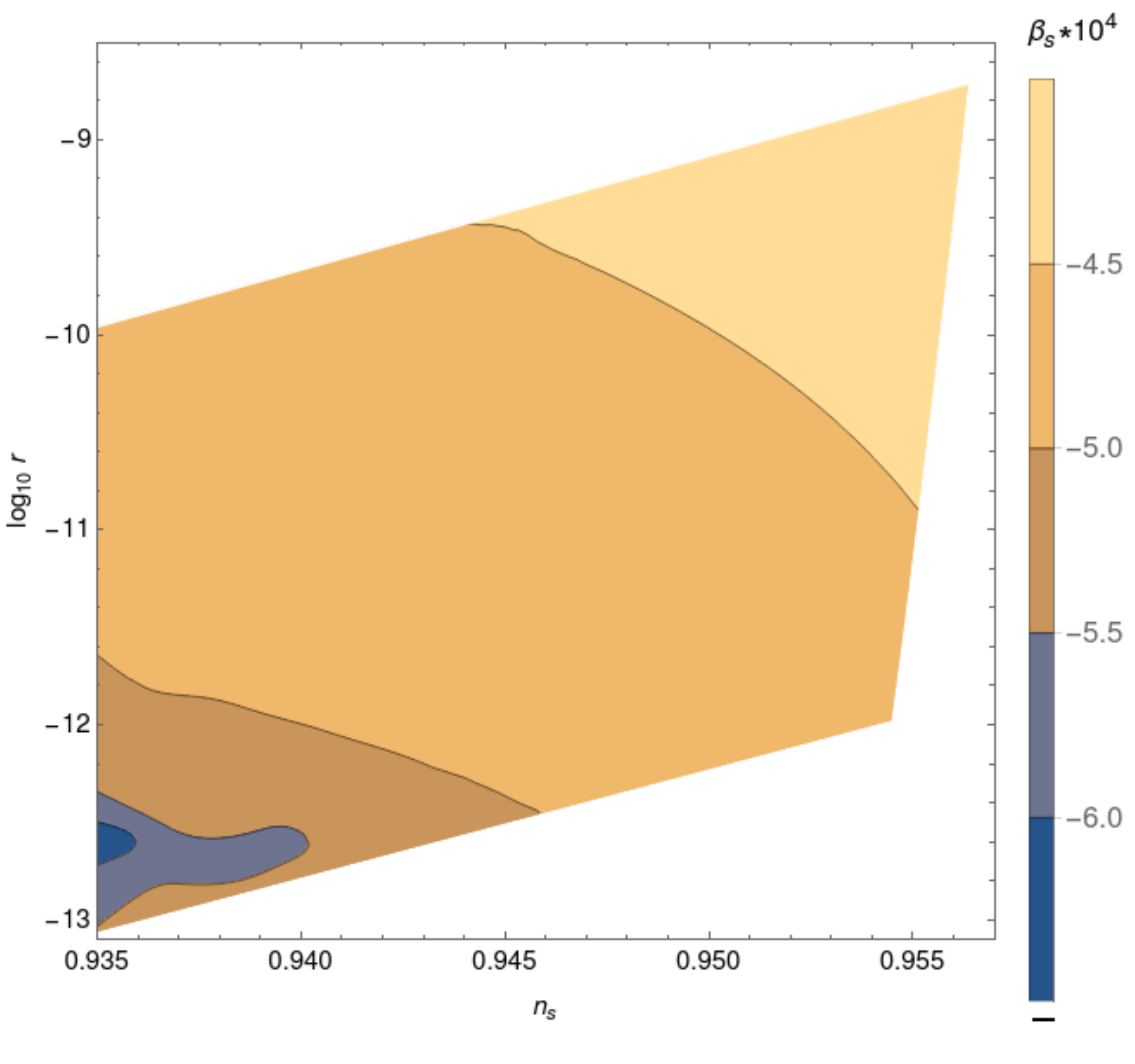}
\end{center}
\caption{Numerical values of $\alpha_s$ and $\beta_s$ on the $(n_s,r)$ plane in the Palatini case.}
\label{fig:large_numeric_pics_palatini_2}
\end{figure}

The results agree well with our analytical approximation, although, as mentioned earlier, the numerical analysis sets stronger restrictions on the allowed parameter space, in particular as regards $r$.

In \cite{Rasanen:2017ivk}, Higgs inflation at hilltop in the Palatini case was considered using an analytical approximation for the potential. Our results are consistent both for the numerical value of $n_s$ and $r$ and for the correlations between the variables (increasing $n_s$ increases $r$ and decreases $\xi$). Our order of magnitude for $\alpha_s$ and $\beta_s$ is the same, but our lower limit for $r$ extends three orders of magnitude below theirs.

As in the metric case, varying the electroweak scale masses $m_H$ and $m_t$ within the 2$\sigma$ interval affects only the second or higher digits of the limits quoted above. The average value of $\Delta\lambda$ is $0.03$ and the average value of $\Delta y_t$ is $0.02$. Putting the jumps to zero and keeping the Higgs mass at the mean value \re{eq:SM_bestfit_vals}, the value $m_t=171.33$ GeV (inside the observational 95\% confidence interval) gives $n_s=0.94$, $r=1.2\times10^{-11}$, $\alpha_s=-1.1\times10^{-3}$ and $\beta_s=-4.8\times10^{-5}$. The value of $n_s$ is just within observational 95\% confidence interval, and the other observables are well below the observational upper limits.

\section{Conclusions} \label{sec:conclusions}

We have performed the first in-depth study of Higgs inflation at the hilltop with renormalization group running, including a detailed scan of the parameter space. With plateau inflation \cite{Fumagalli:2016lls, Enckell:2016xse} and critical point inflation \cite{Allison:2013uaa, Bezrukov:2014bra, Hamada:2014iga, Bezrukov:2014ipa, Rubio:2015zia, Fumagalli:2016lls, Enckell:2016xse, Rasanen:2017ivk} having already been studied in detail, the hilltop scenario is the third case to be scrutinized by careful renormalization group analysis, with hillclimbing inflation \cite{Jinno:2017a, Jinno:2017b} (which also relies on quantum corrections) remaining to be similarly mapped.

At small field values $h\ll1/\xi$, we have calculated the effective Higgs potential using the SM renormalization group equations with the two-loop effective potential and three-loop beta functions, while for large field values $h\gg1/\sqrt{\xi}$ we use the one-loop chiral SM effective potential and one-loop beta functions. The renormalization scale is chosen to minimize the chiral SM one-loop correction at the hilltop.  We require that a hilltop is formed at given values of $\delta\equiv1/(\xi h^2)$ and the Higgs self-coupling $\lambda$, and proceed to solve the whole potential from these inputs. We run the equations up from the electroweak scale and down from the hilltop, and match in the intermediate regime, allowing for an arbitrary jump in the couplings $\lambda$ and $y_t$. We calculated the values of the CMB observables both in the metric and the Palatini formulation of general relativity.

We have confirmed that without the non-minimal coupling, a hilltop inflation in agreement with observations is not possible. In the intermediate regime $\delta\sim1$, successful inflation is not possible either. However, if the hilltop is formed at large field values, $\delta\ll1$, the results agree with observations.

For the metric case, we obtain $n_s\leq0.96$. When $n_s$ is restricted to the observational 95\% confidence interval \eqref{eq:sr_obs_n}, we have $1.3\times10^{-4}<r<1.2\times10^{-3}$, $-0.93\times10^{-3}<\alpha_s<-0.76\times10^{-3}$ and $-3.8\times10^{-5}<\beta_s<-3.1\times10^{-5}$. The result for $n_s$ is at most equal to the value for tree-level plateau inflation ($n_s=0.96$) \cite{Bezrukov:2007ep}, whereas $r$ is always smaller than the plateau counterpart ($r=4.8\times10^{-3}$). Even if we match the couplings run up from the electroweak scale and down from the hilltop exactly without any jumps, we find (for $m_H=125.53$ GeV and $m_t=171.29$ GeV) results that agree well with observations, with $n_s=0.96$, $r=1.1\times10^{-3}$, $\alpha_s=-0.81\times10^{-3}$ and $\beta_s=-3.2\times10^{-5}$.

In the Palatini case, we again have $n_s\le0.96$, When $n_s$ is restricted to the observational 95\% confidence interval, we get $2.2\times10^{-13}<r<2.2\times10^{-9}$, $-1.0\times10^{-3}<\alpha_s<-0.94\times10^{-3}$ and $-4.8\times10^{-5}<\beta_s<-4.0\times10^{-5}$. If we put the jumps to zero, we now find (for the mean value of the Higgs mass and $m_t=171.33$ GeV) the values $n_s=0.94$, $r=1.2\times10^{-11}$, $\alpha_s=-1.1\times10^{-3}$ and $\beta_s=-4.8\times10^{-5}$. In the Palatini formulation, the hilltop results don't significantly differ from the tree-level plateau results \cite{Bauer:2008zj}. In particular, $r$ is highly suppressed compared to the metric case, as in the plateau scenario.

In both cases, the numerical calculation is not sensitive to variation of the Higgs and top masses, changing only in the second or higher digit. We have also done an analytical calculation to leading order in $\delta$, finding agreement with the numerical results in both the metric and the Palatini formulation.

These results come with the caveat that we have neglected differences in renormalization group running between the metric and the Palatini formulation \cite{Markkanen:2017tun}. The initial stages of preheating, when the field value is still large, may also be different in the two formulations, though once the field oscillates around the electroweak vacuum, the effect of the non-minimal coupling, and hence the difference between the two formulations, is small. Also, we did not allow jumps in the gauge couplings, like we did for $\lambda$ and $y_t$.

The predictions of Higgs inflation depend both on quantum corrections and on the choice of the gravitational degrees of freedom (in our case, whether the connection is an independent variable). The next generation of CMB experiments such as COrE\footnote{\url{http://www.core-mission.org/}}, LiteBIRD\footnote{\url{http://litebird.jp/eng/}} and PIXIE\footnote{\url{https://asd.gsfc.nasa.gov/pixie/}} are expected to probe down to $r=10^{-3}$. This would allow them to detect gravitational waves from tree-level Higgs inflation on the plateau or at the critical point in the metric case, and thus rule out the Palatini formulation of general relativity (assuming that Higgs is the inflaton and there are no new degrees of freedom up to the inflationary scale). However, according to our results, even in the metric case, hilltop inflation allows $r$ to lie below the the observational reach of the next generation experiments. So while they will be able to tell the difference between plateau, critical point and hilltop inflation in the metric formulation, they cannot distinguish between the metric and the Palatini formulation at the hilltop. Of course, a more precise determination of $n_s$, or a detection of $\alpha_s$ or $\beta_s$, could rule out Higgs inflation, independent of the value of $r$.

\acknowledgments

SR thanks Fedor Bezrukov for correspondence. ET is supported by the Vilho, Yrj{\"o} and Kalle V{\"a}is{\"a}l{\"a} Foundation of the Finnish Academy of Science and Letters.

%\section*{References}
\bibliographystyle{JHEP}
\bibliography{hh}

%\printbibliography

\end{document}